\documentclass[a4paper,12pt]{article}
\usepackage[utf8x]{inputenc}
\usepackage[left=2cm,right=2cm,top=2cm,bottom=3cm]{geometry}

\usepackage[linktoc=all,hidelinks]{hyperref}
\usepackage{cite}
\usepackage{calrsfs}
\usepackage{amsmath}
\usepackage{amsfonts}
\usepackage{amssymb}
\usepackage{appendix}
\usepackage{caption}
\usepackage{xcolor}

\usepackage{tensor}
\usepackage[smalltableaux,centertableaux,centerboxes]{ytableau}
\usepackage{dynkin-diagrams}
\newcommand{\pd}{\partial}

\makeatletter
\newcommand{\doublewidetilde}[1]{{%
  \mathpalette\double@widetilde{#1}%
}}
\newcommand{\double@widetilde}[2]{%
  \sbox\z@{$\m@th#1\widetilde{#2}$}%
  \ht\z@=0.88\ht\z@
  \skew{1.5}\widetilde{\box\z@}%
}
\makeatother

\begin{document}
\numberwithin{equation}{section}

\thispagestyle{empty}
\begin{center}

\vspace*{50pt}
{\LARGE \bf Higher dualisations of linearised gravity\\ and the $A_1^{+++}$ algebra}

\vspace{30pt}
Nicolas Boulanger${}^{\,a}$,
Paul P. Cook${}^{\,b}$,
Josh A. O'Connor${}^{\,a}$
and Peter West${}^{\,b}$

\vspace{15pt}
\centering
\href{mailto:nicolas.boulanger@umons.ac.be}{\texttt{nicolas.boulanger@umons.ac.be}}
\quad
\href{mailto:paul.cook@kcl.ac.uk}{\texttt{paul.cook@kcl.ac.uk}}
\\
\href{mailto:josh.o'connor@umons.ac.be}{\texttt{josh.o'connor@umons.ac.be}}
\quad
\href{mailto:peter.west540@gmail.com}{\texttt{peter.west540@gmail.com}}

\vspace{15pt}
\centering${}^a$ {\sl \small
Physique de l’Univers, Champs et Gravitation, Université de Mons\quad\null\\
Place du Parc 20, 7000 Mons, Belgium\quad\null}\\
\vspace{10pt}
\centering${}^b$ {\sl \small
Department of Mathematics, King’s College London\quad\null\\
Strand, London, WC2R 2LS, UK\quad\null}\\

\vspace{50pt}
{\bf{Abstract}} 
\end{center}

\noindent The non-linear realisation based on $A_1^{+++}$ is known to describe gravity in 
terms of both the graviton and the dual graviton. We extend this analysis 
at the linearised level to find the equations of motion for the first higher dual 
description of gravity that it contains. We also give a systematic method for 
finding the additional fields beyond those in the non-linear realisation that
are required to construct actions for all of the possible dual descriptions
of gravity in the non-linear realisation. We show that these additional fields 
are closely correlated with the second fundamental representation 
of $A_1^{+++}\,$.

\newpage

\setcounter{tocdepth}{2}
\tableofcontents

\newpage


\section{Introduction}

It was shown that the conjectured \cite{West:2001as,West:2003fc} 
non-linear realisation of the semi-direct product $E_{11}\ltimes\ell_1$
of $E_{11}$ with its vector representation contains the
fields and the equations of motion of every 
maximal supergravity theory \cite{Tumanov:2015yjd,Tumanov:2016abm}. 
For a review, see \cite{West:2016xro}. 
As such, it contains the metric of gravity and the three form in 
eleven dimensions and there are very good reasons to believe that these are the only 
degrees of freedom that the non-linear realisation possesses 
\cite{West:2019gyl,Glennon:2021ane}. 
However, the non-linear realisation contains an infinite number of fields, of which
only a few are the usual fields of the maximal supergravity theories. 

It was conjectured in \cite{Riccioni:2006az} and 
proven in \cite{Boulanger:2012df} that many of these remaining 
fields represent equivalent descriptions of the degrees
of freedom of the maximal supergravity theories.
For example, in $E_{11}$ at levels $1,4,7,\ldots\,$, we find the
fields $A_{a_1a_2a_3}\,$, $A_{a_1\ldots a_9,b_1b_2b_3}\,$,  
$A_{a_1\ldots a_9, b_1\ldots b_9, c_1c_2c_3}\,$, and so on,
which are related by an infinite set of duality relations.
This ensures that the only degrees of freedom are those
which are usually contained in the first field, the three form. 
However, any of these fields can be used to give an equivalent
formulation of these degrees of freedom.  
At levels $2,5,8,\ldots$\,, the story is similar except that the
the block of three indices $a_1a_2a_3$ in each field is replaced
by a block of six indices $a_1\ldots a_6\,$. 
Then, at levels $0,3,6,9,\ldots\,$, we find fields associated 
with gravity. Indeed, at level zero, we find the usual description
of gravity with the field $h_{ab}\,$.
At level three, we find the field $A_{a_1\ldots a_8,b}$ which was
proposed to provide a dual description of gravity, while at level six
we have $A_{a_1\ldots a_9, b_1\ldots b_8, c}$\,, at level nine we find  
$A_{a_1\ldots a_9, b_1\ldots b_9,  c_1\ldots c_8,d},\ldots$\,, and so on.
These fields also provide alternative descriptions of gravity and all
the fields are related by a set of duality relations which
ensure that the theory only propagates a single graviton.
In fact, there are other fields in the non-linear realisation
and some of these are required to account for the gauged supergravities. 

It is useful to give an account of the history of the dual graviton field.
It was first observed by Curtright that the field
$A_{a_1a_2,b}=A_{[a_1a_2],b}$ could describe pure gravity in five
dimensions \cite{Curtright:1980yk}. 
It was then proposed that the field $A_{a_1\ldots a_{D-3}, b}$ may
describe pure gravity in $D$ dimensions \cite{Hull:2000zn}. 
In order the show that the field $A_{a_1\ldots a_8, b}$ at level three
in the non-linear realisation of $E_{11}\ltimes\ell_1$ did indeed
describe gravity, a parent action in $D$ dimensions was given in
\cite{West:2001as,West:2002jj}. 
By first linearising the parent action of \cite{West:2001as},
then varying the result with respect to one field or the other
and finally substituting inside the linearised parent action,
we obtain either the Fierz-Pauli action in the form where
local Lorentz invariance holds, i.e. in terms of the field $h_{ab}$
that is neither symmetric nor antisymmetric, or we obtain an
action in terms of the dual field $A_{a_1\ldots a_{D-3},b}$\,.
This result was fully explained and also extended to higher
spin fields in reference \cite{Boulanger:2003vs}.
As shown in \cite{West:2001as,West:2002jj}, the parent action 
of \cite{West:2001as}
also led to duality relations between the two fields.
In this way, it was clear that the dual graviton field
$A_{a_1\ldots a_{D-3}, b}$ really did provide an equivalent formulation
of gravity at the linearised level. 
Further connections were also established in \cite{Boulanger:2003vs}
between \cite{West:2001as}, \cite{Curtright:1980yk} and \cite{Hull:2000zn}. 
These developments are reviewed at the beginning of
Section \ref{section:off-shell}.  

It was also conjectured that the non-linear realisation of 
the semi-direct product $A_{D-3}^{+++}\ltimes\ell_1\,$ 
of the very-extended algebra $A_{D-3}^{+++}$ with its vector representation,
contains pure gravity in $D$ dimensions \cite{Lambert:2006he}. 
Following early preparatory work in references \cite{Tumanov:2014pfa} 
and \cite{Pettit:2017zgx}, this was indeed shown to be the case in four 
and eleven dimensions \cite{Glennon:2020qpt} and \cite{Glennon:2020uov} 
respectively. In four dimensions, at the lowest level, this non-linear 
realisation contains the usual field of gravity $h_{ab}$\,.
At higher levels -- indicated by numbers in brackets after each field -- 
in addition to other fields, it contains 
\begin{align}
    h_a{}^b\,(0)\,; \quad A_{(ab)}\,(1)\,; \quad A_{a_1a_2,(b_1b_2)}\,(2)\,; \quad A_{a_1 a_2,b_1 b_2,(c_1 c_2)}\,(3)\,; \quad A_{a_1 a_2,b_1 b_2, c_1 c_2,(d_1 d_2)}\,(4)\,; \quad \ldots 
\label{eq(1.1)}
\end{align}
where groups of indices are antisymmetric unless otherwise indicated by round
brackets $(\,\cdots)$ in which case they are symmetric.
We interpret these fields as being related to dual descriptions of gravity.
The field at level one is called the dual graviton. We then find the first
higher dual graviton at level two, the second higher dual graviton at level
three, and so on. The equations of motion at the full non-linear level,
as well as the duality relations, were found for $h_a{}^b$ and $A_{(ab)}$
in four dimensions \cite{Glennon:2020qpt}
and in eleven dimensions \cite{Glennon:2020uov}.

The non-linear realisations of $E_{11}\ltimes\ell_1$ and
$A_{D-3}^{+++}\ltimes\ell_1$ lead to an infinite number of duality
relations which can then be used to derive the equations of motion
of the fields. These field equations are constructed from fields 
that are irreducible representations of $A_{D-1}$
and, as a result,
they have more and more space-time derivatives for the fields at
higher and higher levels. The equations of motion require only
the fields in the non-linear realisation, and they correctly
describe the relevant degrees of freedom.
In \cite{Bekaert:2002dt} and \cite{Bekaert:2003az}, equations
of this type which describe the irreducible representations 
of the Poincar\'e group were given precisely.
As shown in \cite{Bekaert:2003az} and reviewed in \cite{Bekaert:2006ix}, 
one can also integrate these equations to find equations of motion
that are second order in space-time derivatives provided that one makes
a particular gauge choice that leads to the Labastida 
\cite{Labastida:1986ft,Labastida:1987kw} gauge transformations
for arbitrary mixed-symmetry fields where the gauge parameters
obey trace constraints. In fact, the duality relations
derived from the non-linear realisation only hold modulo gauge
transformations which can, as a matter of principle, be deduced from
the non-linear realisation. See, for example, \cite{Tumanov:2016abm}
or the review \cite{West:2017mqg}. However, one must also introduce
extra fields in order to have duality relations that hold as equations
of motion in the usual sense and not just as equivalence
relations \cite{Boulanger:2015mka}.

A parent action containing the fields  $A_{a_1a_2a_3}\,$, 
$A_{a_1\ldots\,a_9, b_1b_2b_3}$\,, $\ldots\,$ occurring at levels $1,4,\ldots$
in the $E_{11}\ltimes\ell_1$ non-linear realisation, also containing
certain extra fields, was worked out in \cite{Boulanger:2015mka}
along the lines of \cite{Boulanger:2012df,Boulanger:2012mq}.
Depending on which field was eliminated, one found an action only in
terms of one field or the other. In this way, the authors 
of \cite{Boulanger:2015mka} found an action for the latter field 
which we call the first higher dual of the three from. 
The higher level fields were also discussed in \cite{Boulanger:2015mka}, 
as were the infinite chain of duality relations and analogous results for 
the six form. Hence, using parent actions, one could find the
additional fields required in order to write down an action, 
or duality relations, for the higher dual fields.

A similar strategy had previously been suggested for pure gravity
in \cite{Boulanger:2012df}. The method of parent actions was used to
produce, for the first time, an infinite number of higher dual action
principles, 
thereby proving the conjecture established in \cite{Riccioni:2006az} 
on the equivalent dual descriptions of gravity. These parent actions
involve extra fields in comparison to those that appear in the non-linear
realisation of $A_1^{+++}\ltimes\ell_1\,$. 

In this paper, we further pursue the approach set forth in
\cite{Boulanger:2012df} 
to higher dual descriptions of gravity, focusing on four space-time 
dimensions for the sake of concreteness. We provide an explicit procedure
for constructing the parent actions that relate the different higher dual
formulations of gravity. Using these parent actions, one can directly obtain
action principles for each subsequent higher dual graviton. We find 
extra fields on top of those already in the non-linear realisation of 
$A_{1}^{+++}\ltimes\ell_1$\,. These extra fields are required to formulate actions
for the dual fields as well as the duality relations between dual fields at
adjacent levels. 

We will compare the type of additional fields required to form higher dual actions 
with those contained in the adjoint representation and the second fundamental
representation, denoted $l_2\,$, of $A_{1}^{+++}\,$.
While there is a striking agreement between the $\mathrm{GL}(4)$-irreducible symmetry
types of the extra fields appearing in the higher dual actions 
and the $\ell_2$ representation of $A_1^{+++}$\,,
the number of times each type of extra field appears off-shell 
does not always coincide with their multiplicities in the $\ell_2$
representation.

\section{The Kac-Moody algebra $A_1^{+++}$}

\subsection{The non-linear realisation of $A_1^{+++}\ltimes\ell_1$}
\label{subsec:non-linear}

Following earlier results \cite{Tumanov:2014pfa,Pettit:2017zgx}, the non-linear
realisation of the semi-direct product $A_1^{+++}\ltimes\ell_1$ of $A_1^{+++}$ with its vector representation was computed at low levels in \cite{Glennon:2020qpt}. This calculation will be reviewed in this section.
The Dynkin diagram for $A_1^{+++}$ is
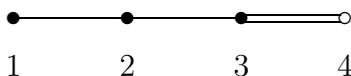
\begin{equation*}
    \begin{matrix}
    \bullet & \hspace{-0.5cm}\frac{\phantom{\hspace{1.5cm}}}{\phantom{\hspace{1.5cm}}}\hspace{-0.5cm} & \bullet & \hspace{-0.5cm}\frac{\phantom{\hspace{1.5cm}}}{\phantom{\hspace{1.5cm}}}\hspace{-0.5cm} & \bullet & \hspace{-0.4cm}\overline{\overline{\hspace{1.25cm}}}\hspace{-0.4cm} & \circ \\
    1 & & 2 & & 3 & & 4
    \end{matrix}
\end{equation*}
While no complete description of the generators of any such Kac-Moody algebra exists,
they can still be analysed by decomposing them with respect to certain subalgebras.
Deleting node 4 from the Dynkin diagram of $A_1^{+++}$ allows us to analyse the
algebra in terms of its decomposition into $\mathrm{GL}(4)$ \cite{Tumanov:2014pfa}.
The resulting generators can be classified in terms of a level which, in this case,
is the number of up minus down $\mathrm{GL}(4)$ indices divided by two. The positive low
level generators $R^{\underline \alpha}$ are given, alongside the
level zero generator, by
\begin{gather}
    K^a{}_b\,(0)\,; \quad R^{(ab)}\,(1)\,; \quad R^{a_1a_2,(b_1b_2)}\,(2)\,; \quad R^{a_1a_2,b_1b_2,(c_1c_2)}\,(3)\,, \quad R^{a_1a_2a_3,b_1b_2,c}\,(3)\,;\nonumber\\
    R^{a_1a_2,b_1b_2,c_1c_2,(d_1d_2)}\,(4)\,, \quad R^{a_1a_2a_3,b_1b_2,(c_1c_2c_3)}\,(4)\,, \quad R^{a_1a_2a_3,b_1b_2,c_1c_2,d}_{(1)}\,(4)\,\ \quad R^{a_1a_2a_3,b_1b_2,c_1c_2,d}_{(2)}\,(4)\,,\nonumber\\
    R^{a_1a_2a_3,b_1b_2b_3,(c_1c_2)}\,(4)\,, \quad R^{a_1a_2a_3a_4,b_1b_2,(c_1c_2)}\,(4)\,, \quad R^{a_1a_2a_3a_4,b_1b_2b_3,c}\,(4)\,; \quad \ldots\label{eq(1.1.1)}
\end{gather}
The number in  brackets corresponds to the level of the generators and the subscripts
enumerate the generators when there is more than one with the same index structure.
Groups of indices are antisymmetric except when shown to be symmetrised using round 
brackets. For example, the generator $R^{ab,cd}$ satisfies
$R^{ab,cd}=R^{[ab],cd}=R^{ab,(cd)}\,$. The generators belong to irreducible
representations of $\mathrm{GL}(4)\,$, i.e they all satisfy 
over-antisymmetrisation irreducibility conditions. 
For example, $R^{[ab,c]d}=R^{[ab|,c|d]}= 0$\,. 
Negative level
generators have the same index structure with lowered indices. 
Commutation relations for these $A_1^{+++}$ generators can be found in \cite{Glennon:2020qpt}. 

The generators in the vector representation of the $A_1^{+++}$ are denoted by $L_A$ and, when decomposed into representations of $\mathrm{GL}(4)$, the low level generators found in \cite{Glennon:2020qpt} are given by
\begin{gather}
    P_a\,(0)\,; \quad Z^a\,(1)\,; \quad Z^{(a_1a_2a_3)}\,(2)\,, \quad Z^{a_1a_2,b}\,(2)\,; \quad Z^{a_1a_2,(b_1b_2b_3)}\,(3)\,,\nonumber\\
    \quad Z^{a_1a_2,b_1b_2,c}_{(1)}\,(3)\,, \quad Z^{a_1a_2,b_1b_2,c}_{(2)}\,(3)\,, \quad Z^{a_1a_2a_3,(b_1b_2)}\,(3)\,, \quad Z^{a_1a_2a_3,b_1b_2}\,(3)\,,\nonumber\\
    \quad Z^{a_1a_2,b_1b_2,(c_1c_2c_3)}_{(1)}\,(4)\,, \quad Z^{a_1a_2,b_1b_2,(c_1c_2c_3)}_{(2)}\,(4)\,, \quad \ldots, 
    \label{eq(1.1.2)}
\end{gather}
where, as before, groups of indices are antisymmetric except for those in round 
brackets which are symmetric. Subscripts denote different generators when the
multiplicity is greater than one.  These generators satisfy the usual
$\mathrm{GL}(4)$-irreducibility conditions. For example, $Z^{[a_1 a_2,b]} = 0$\,.
Generators in the vector representation commute and their commutators with the generators of $A_1^{+++}$ are given in \cite{Glennon:2020qpt}. 

The construction of the equations of motion follows the same pattern as that for
{$E_{11}\ltimes\ell_1$}\,. See \cite{West:2016xro,West:2017mqg} for reviews. For the non-linear realisation based on 
{$A_1^{+++}\ltimes\ell_1$} in \cite{Glennon:2020qpt}, we start from 
the group element of {$A_1^{+++}\ltimes\ell_1$} denoted by 
$g=g_Lg_A$\,, where $g_A$ and $g_L$ are group elements that are constructed in
terms of non-negative level generators of the adjoint and vector representations,
respectively, of $A_1^{+++}$\,. They take the form 
\begin{align}
    g_A=e^{A_{\underline \alpha} R^{\underline \alpha}} := {}&\cdots \exp(A_{ab,cd}\,R^{ab,cd})\,\exp(A_{ab}\,R^{ab})\,\exp(h_{a}{}^{b}\,K^{a}{}_{b})\;,\label{eq(1.1.3)}\\
    g_L=e^{z^AL_A} := {}&\exp(x^a P_a)\exp(z_aZ^a)\exp(z_{abc}Z^{abc}+z_{ab,c}Z^{ab,c})\cdots\;.\label{eq(1.1.4)}
\end{align}
Therefore, the theory is populated by a set of fields $A_{\underline \alpha}$
which contains the graviton $h_a{}^b$\,, the dual graviton $A_{ab}$\,, the first 
higher dual graviton $A_{ab,cd}$\,, and so on. We see from the list of generators
in \eqref{eq(1.1.1)} that we have the generator
$R^{a_1 a_2,b_1 b_2, c_1 c_2,(d_1 d_2)}$ at level four which results in the second
higher dual graviton $A_{a_1 a_2,b_1 b_2, c_1 c_2,(d_1 d_2)}$\,.
Indeed, the pattern continues so that one finds such fields at every level.
This leads to an infinite tower of dual formulations of pure gravity with fields
that depend on the generalised coordinates 
$z^A=\left\{ x^a, z_a, z_{abc}, z_{ab,c}, \ldots \right\}$\,. 

The non-linear realisation is invariant under rigid transformations
{$g_0\in{}A_1^{+++}\ltimes\ell_1$} and local transformations
$h\in{}I_c(A_1^{+++})$\,, where $I_c(A_1^{+++})$ is the Cartan involution
invariant subalgebra of $A_1^{+++}$\,. This means that generic group elements 
$g=g_Lg_A$ are invariant under
\begin{equation}
    g \to g_0 g \qquad\quad \mathrm{and} \quad\qquad g \to g h\;,
\label{eq(1.1.5)}
\end{equation}
where $g_0$ is a rigid (i.e. constant) group element and $h$ is a local
transformation which can be used to set the coefficients of all negative level
generators in $g_A$ to zero \cite{West:2014qoa}. The equations of motion are just
those that are invariant under these transformations and, as for $E_{11}$\,, they
are essentially unique. 

The dynamics of the non-linear realisation is often constructed 
using Maurer-Cartan forms
\begin{equation}
    \nu \equiv g^{-1} \mathrm{d} g = \nu_A + \nu_L\;,
    \label{eq(1.1.6)}
\end{equation}
where $\nu_A=g_A^{-1}\,\mathrm{d}g_A \equiv \mathrm{d}z^\Pi\,G_{\Pi,\underline\alpha}\,R^{\underline \alpha}$
and $\nu_L= g_A^{-1}(g_L^{-1}\mathrm{d}g_L) g_A = g_A^{-1}\,(\mathrm{d}z \cdot L)\, g_A \equiv
\mathrm{d}z^\Pi E_\Pi{}^A L_A $\,.
Here, ${E}_{\Pi}{}^A$ can be thought of as a vierbein on the generalised
space-time. Its lowest component is the gravitational vierbein given by
${e_\mu}^a={(\exp(h))_\mu}^a$. The $G_{\Pi,\underline{\alpha}}$ are the
components of the Maurer-Cartan form where the index $\Pi$ is a world-volume 
(derivative) index and $\underline \alpha$ is an index in the adjoint
representation.  

The low level Maurer-Cartan forms in the $A_1^{+++}$ direction are given by
\begin{gather}
    G_{\Pi,}{}_a{}^b={}e_a{}^\mu \partial_\Pi e_\mu {}^b\;,
    \qquad\qquad
    \overline{G}_{\Pi,bc}={e_b}^\kappa {e_c}^\lambda \partial_\Pi A_{\kappa \lambda}\;,\\
    {{\overline{\overline {G}}}}_{\Pi,a_1a_2,bc}
    ={e_{a_1}}^{\kappa_1}{e_{a_2}}^{\kappa_2} {e_b}^{\lambda_1} {e_c}^{\lambda_2} \left(\partial_\Pi A_{\kappa_1 \kappa_2,\lambda_1\lambda_2}-A_{[\kappa_1 |(\lambda_1}\partial_\Pi A_{\lambda_2)|\kappa_2]}\right)\;.
    \label{eq(1.1.7)}
\end{gather}
They are found as the coefficients of $K^a{}_b$\,, $R^{bc}$ and $R^{a_1a_2,bc}$
in the Maurer-Cartan form, where $\partial_\Pi$ is the derivative with respect to 
the coordinates $z^\Pi$\,. The dynamics is actually constructed from 
$G_A{}_{, \underline \alpha}\equiv ( E^{-1})_A{}^\Pi G_\Pi{}_{, \underline \alpha}= e_a{}^\mu G_\mu{}_{, \underline \alpha}+\cdots$\,,
where ``$\cdots$'' corresponds to
terms arising when higher contributions to the vierbein
${E}_{\Pi}{}^A$ are taken into account. These contributions contain derivatives
with respect to the higher level coordinates. Working with
$G_A{}_{, \underline \alpha}$ has the advantage that it only transforms
under $I_c(A_1^{+++})$\,.

Rather than deriving the equations of motion, one can derive a set of duality
relations from which the equations of motion can be deduced, as explained
in \cite{West:2016xro,West:2017mqg,Tumanov:2015yjd}.
The duality relations between low level fields at the lowest level of 
generalised space-time derivatives are \cite{Glennon:2020qpt}
\begin{align}
    E_{a,b_1b_2}\equiv{}&(\det e)^{1 / 2}\omega_{a,b_1b_2}+\tfrac{1}{2}\, {\varepsilon_{b_1b_2}}^{c_1c_2}\overline{G}_{c_1,c_2a} \;\dot{=}\; 0\;,\label{eq(1.1.8)}\\
    \overline{E}_{a,b_1b_2}\equiv{}&\overline{G}_{a,b_1 b_2} + \varepsilon_a{}^{c_1 c_2 c_3} {{\overline{\overline {G}}}}_{c_1 , c_2 c_3 , b_1 b_2} \;\dot{=}\; 0\;,\label{eq(1.1.9)}
\end{align}
where $\omega_{a,b_1b_2}$ is the usual expression for the spin connection in terms 
of the vierbein which is given in terms of the low level Maurer-Cartan forms by
\begin{equation}
    (\det e)^{1/2} \omega_{a,b_1b_2}=-\,G_{b_1,(b_2a)}+G_{b_2,(b_1a)}+G_{a,[b_1b_2]}
    \label{eq(1.1.10)}
\end{equation}
with $G_{a,bc}=e_a{}^\mu {e_b}^\nu \partial_\mu e_{\nu c}$\,.

Equation \eqref{eq(1.1.8)} relates the graviton field $h_a{}^b$ appearing at level
zero to the dual graviton field $A_{ab}$ at level one, while equation \eqref{eq(1.1.9)} is a
duality relation between $A_{ab}$ at level one and the first higher dual graviton
field $A_{a_1a_2,bc}$ appearing at level two.

By combining equations \eqref{eq(1.1.8)} and \eqref{eq(1.1.9)}, one derives a
duality relation between the graviton and the first higher dual graviton:
\begin{equation}
    E'_{a,b_1b_2} \equiv (\det e)^{1 / 2}\omega_{a,b_1b_2}+3\,
    \overline{\overline {G}}{}_{[b_1,b_2 c]}{}^c{}_{,a} \;\dot{=}\; 0\;.\label{eq(1.1.11)}
\end{equation}

The above duality relations only hold modulo certain gauge transformations,
as indicated by the symbol ``$\,\dot{=}\,$'', so
they really are equivalence relations. This is explained in 
\cite{West:2017mqg,Tumanov:2016abm,Tumanov:2015yjd,West:2014qoa}.
In order to further manipulate the above duality relations, one needs to know what
the gauge transformations are. We will obtain them in the next section.
As explained in \cite{West:2001as}, the duality relation \eqref{eq(1.1.8)}
may be turned into a usual equation by adding an antisymmetric component
to the symmetric dual graviton $A_{a_1a_2}$\,.
This 2-form field will later be found inside the second fundamental representation of $A_1^{+++}$\,.
In what follows we are only concerned with the linearised theory and so we drop,
in particular, the $\det e$ factors.

\subsection{Gauge transformations}\label{subsection:A1+++gaugetransfo}

It was proposed in \cite{West:2014eza} that a theory constructed from a non-linear
realisation of $\mathfrak{g}^{+++}\ltimes\ell_1$\,, where $\mathfrak{g}^{+++}$ is
any very-extended Kac-Moody algebra and $\ell_1$ is its vector (first fundamental)
representation, is invariant under a particular set of gauge transformations whose 
parameters are in a one-to-one correspondence with the spectrum of $\ell_1$\,. For the linearised theory where base and fiber indices are 
identified, these gauge transformations take the form
\begin{equation}
    \delta A_{\underline{\alpha}}=C^{-1}_{\underline{\alpha},\underline{\beta}}(D^{\underline \beta})_E{}^F\partial_F \Lambda^E\;. \label{eq(1.2.1)}
\end{equation}
In this equation, $C^{-1}_{\underline{\alpha},\underline{\beta}}$ is the inverse
of the Cartan-Killing metric $C^{\underline{\alpha},\underline{\beta}}$ for
$\mathfrak{g}^{+++}$. The matrix $(D^{\underline \beta})_E{}^F$ is that for the
vector representation and, in particular, it occurs in the commutator
\begin{equation}
    [R^{\underline{\alpha}},L_A] = - (D^{\underline{\alpha}})_A{}^B L_B\;.
    \label{eq(1.2.2)}
\end{equation}
In addition, we have used the partial derivative in the linearised theory
$\partial_F= \tfrac{\partial}{\partial z^F}$\,.
The gauge parameters $\Lambda^A$
correspond to elements in the vector representation.

Hence, in order to evaluate the gauge transformations, we require the inverse
Cartan-Killing matrix and the analogous matrix for $\ell_1$
at the corresponding level.
The gauge transformations for the graviton and the dual graviton in the non-linear
realisation of $A_1^{+++}\ltimes\ell_1$ were computed in \cite{Glennon:2020qpt} and
we now extend these previous results to the main object of study in this paper:
the first higher dual graviton.

We begin with the computation of the Cartan-Killing form which is determined by
requiring that it is invariant. For our current purposes this means that it should
satisfy
\begin{equation}
    ([R^{a_1a_2}, R^{b_1b_2} ] , R_{c_1c_2, d_1d_2} )+ (R^{a_1a_2},    [R_{c_1c_2, d_1d_2}, R^{b_1b_2}] )=0\;,
    \label{eq(1.2.3)}
\end{equation}
where $(\cdot\,,\cdot)$ is the symmetric non-degenerate bilinear form on $A_1^{+++}$
that generalises the Killing form for finite-dimensional semi-simple Lie algebras. One finds that
\begin{equation}
    C^{a_1a_2 , b_1b_2,} {}_{c_1c_2, d_1d_2} = \delta ^{a_1a_2}_{c_1c_2}\delta ^{b_1b_2} _{( d_1d_2)} + \delta^{[ a_1 | b_1} _{c_1c_2}\delta ^{| a_2 ] b_2}_{( d_1d_2 )}+ \delta^{[ a_1 | b_2} _{c_1c_2}\delta ^{| a_2 ] b_1}_{( d_1d_2 )}
    \label{eq(1.2.4)}\;,
\end{equation}
where $\delta^{a_1a_2}_{(b_1b_2)}= \delta^{a_1 }_{(b_1}\delta ^{a_2}_{b_2)}$ in
contrast to the usual symbol 
$\delta^{a_1a_2}_{b_1b_2}= \delta^{a_1 }_{[b_1}\delta ^{a_2}_{b_2]}$\,.

Taking the previous results from \cite{Glennon:2020qpt}, we find that the
Cartan-Killing metric up to the level of the first higher dual graviton is
given by
\begin{equation}
    C_{\underline{\alpha},\underline{\beta}} = 
    \left(
    \begin{matrix}
    \delta^c_b \delta^a_d -\tfrac{1}{2}\delta^a_b \delta^c_d & 0 & 0 & 0 & 0 \cr
    0 & 0 & \delta^{(a_1a_2)}_{b_1 b_2} & 0 & 0\cr
    0 & \delta^{(a_1a_2)}_{b_1b_2} & 0 & 0 & 0\cr
    0 & 0 & 0 & 0 & C^{a_1a_2 , b_1b_2,} {}_{c_1c_2, d_1d_2} \cr
    0 & 0 & 0 &  C^{c_1c_2 , d_1d_2,} {}_{a_1a_2, b_1b_2}  & 0\cr
    \end{matrix}
    \right)\;.
    \label{eq(1.2.5)}
\end{equation}
where the basis is ordered to match the scalar products of 
$K^a{}_b$\,, $R^{a_1 a_2}$\,, $R_{a_1a_2}$\,, $R^{a_1a_2,b_1b_2}$ and
$R_{a_1a_2,b_1b_2}$ with $K^c{}_d$\,, $R^{b_1 b_2}$\,, $R_{b_1b_2}$\,,
$R^{c_1c_2,d_1d_2}$ and $R_{c_1c_2,d_1d_2}$\,. Note that the only non-zero
entries of $C_{{\underline \alpha},{\underline \beta}}$ are found when the 
levels of $\underline{\alpha}$ and $\underline{\beta}$ sum to zero.

The inverse Cartan-Killing metric is given by
\begin{equation}
    C^{-1}_{\underline{\alpha},\underline{\beta}} = 
    \left(
    \begin{matrix}
    \delta^e_c \delta^d_f -\tfrac{1}{2}\delta^d_c \delta^e_f & 0 & 0 & 0 & 0 \cr
    0 & 0 & \delta^{b_1b_2}_{(e_1e_2)} & 0 & 0\cr
    0 & \delta^{b_1b_2}_{(e_1 e_2)} & 0 & 0 & 0\cr
    0 & 0 & 0 & 0 & C^{e_1e_2 , f_1f_2,} {}_{c_1c_2, d_1d_2} \cr
    0 & 0 & 0 &C^{c_1c_2 , d_1d_2,} {}_{e_1e_2, f_1f_2} & 0\cr
    \end{matrix}
    \right)\;.
    \label{eq(1.2.6)}
\end{equation}

The vector representation appears in the commutators of \eqref{eq(1.2.2)}
which were given in \cite{Glennon:2020qpt} at low levels. Omitting the
commutators with $K^a{}_b$\,, they are given by
\begin{gather}
    [R^{ab},\,P_c] = \delta^{(a}_{\,c}\,Z^{b)}\;, \qquad\quad [R^{ab},\,Z^{c}] = Z^{abc} + Z^{c(a,b)}\;,\nonumber\\
    [R^{ab,cd},\,P_e] = -\,\delta^{[a}_{\,e}\,Z^{b]cd} + \tfrac{1}{4}\left( \delta^a_e\,Z^{b(c,d)} - \delta^b_e\,Z^{a(c,d)} \right) - \tfrac{3}{8}\left( \delta^c_e\,Z^{ab,d} + \delta^d_e\,Z^{ab,c} \right)\;,\nonumber\\
    [R_{ab},\,P_c] = 0\;, \qquad\quad [R_{ab},\,Z^c] = 2\,\delta^{\,c}_{(a}\,P_{b)}\;,\nonumber\\
    [R_{ab},\,Z^{cde}] = \tfrac{2}{3}\left( \delta^{cd}_{(ab)}\,Z^e + \delta^{de}_{(ab)}\,Z^c + \delta^{ec}_{(ab)}\,Z^d \right)\;,\nonumber\\
    [R_{ab},\,Z^{cd,e}] = \tfrac{4}{3}\left( \delta^{de}_{(ab)}\,Z^c - \delta^{ce}_{(ab)}\,Z^d \right)\;.
    \label{eq(1.2.7)}
\end{gather}

To study the vector representation at the level of the
first higher dual graviton, we need to compute certain commutators at higher levels.
One finds that
\begin{gather}
    [R_{a_1a_2, b_1b_2} , Z^{c_1c_2c_3} ]= e_1\,\delta^{(c_1c_2c_3)} _{b_1b_2 [a_1} P_{a_2]}\;, \quad e_1=-4\;,
    \label{eq(1.2.8)}\\
    [R_{a_1a_2, b_1b_2} ,Z^{c_1c_2, d} ]= e_2 \left(\delta^{c_1c_2}_{a_1a_2}\delta^d_{(b_1}P_{b_2)}
    +\delta^{c_1c_2}_{a_1( b_1 |}\delta^d_{a_2}P_{| b_2)}+\delta^{c_1c_2}_{a_1( b_1}\delta^d_{b_2 )}P_{a_2}\right)\;,\quad e_2=2\;,
    \label{eq(1.2.9)}
\end{gather}
where the last expression should be taken so that is is anti-symmetric in $a_1$ and $a_2$.

Using the inverse Cartan-Killing metric \eqref{eq(1.2.6)} and reading off the
analogous matrix for $\ell_1$ from equations
\eqref{eq(1.2.7)}--\eqref{eq(1.2.9)}, we find that the gauge transformations
with gauge parameters
\begin{equation}\label{l1:gaugeparams}
    \Lambda^A:=\{\xi^a\,, \overline{\xi}_a\,, \Lambda_{abc}\,, \Lambda_{ab,c}\,, \ldots\}
\end{equation}
are given, for the fields at low levels, by
\begin{gather}
    \delta h_a{}^b= \partial_a \xi^b\;, \qquad\quad \delta A_{ab}=-2\,\partial_{(a} {\overline \xi}_{b)}\;,
    \label{eq(1.2.10)}\\
    \delta A_{a_1a_2,b_1b_2}= -\,e_1\,\partial_{[a_1}\Lambda_{a_2]b_1b_2}-3\,e_2\, \partial_{(b_1|}  \Lambda_{a_1a_2,|b_2)}
    +2\,e_2\,\partial_{[a_1}\Lambda_{a_2](b_1,b_2)}\;.
    \label{eq(1.2.11)}
\end{gather}
The parameters satisfy $\Lambda_{a_1a_2a_3}=\Lambda_{(a_1a_2a_3)}$ and
$\Lambda_{a_1a_2,b}=\Lambda_{[a_1a_2],b}$ with the irreducibility
condition $\Lambda_{[ a_1a_2, b]}=0$. In these equations, we have not written
the gauge transformations that involve derivatives with respect to the higher
level coordinates. 

\subsection{Linearised equations of motion}
\label{section:a1+++EOM}

The duality relations given in \eqref{eq(1.1.8)}--\eqref{eq(1.1.11)} only hold
modulo certain transformations which arise from the gauge transformations for
the fields involved in the duality relations. As we computed these in the
previous section, we can now compute the resulting transformations up to
which the duality relations hold. Having done this, we can then compute
the equations of motion from the duality relations at the linearised level. 

We first consider the duality relation between gravity and dual gravity in
\eqref{eq(1.1.8)}. Using the gauge transformation \eqref{eq(1.2.11)} we find
that, at the linearised level, it takes the form
\begin{equation}
    E_{a,b_1b_2} \equiv \omega_{a,b_1b_2}+\tfrac{1}{2}\,{\varepsilon_{b_1b_2}}^{c_1c_2}\,\overline{G}_{c_1,c_2a}+\partial_a \xi_{b_1b_2}= 0\;, 
    \label{eq(1.4.1)}
\end{equation}
where
\begin{equation}
    \xi_{b_1b_2}:=-\,\partial_{[b_1}\xi_{b_2]}- {\varepsilon_{b_1b_2}}^{c_1c_2}\,\partial_{c_1}\overline{\xi}_{c_2}\;.
    \label{eq(1.4.2)}
\end{equation}
In deriving this result, we have used local Lorentz symmetry to symmetrise
the $h_{ab}$ field, and so we obtain the variation
$\delta h_{ab} =\partial_{(a}\xi_{b)}\,$. Had we not done this Lorentz gauge
fixing, then the first term on the right-hand-side of \eqref{eq(1.4.2)} would
have been replaced by a Lorentz transformation. We have removed the dot above
the equals sign in \eqref{eq(1.4.1)} since it holds as a usual equation. 

To find the equations of motion, we have to eliminate the gauge
transformations from the duality relations by taking derivatives and, at the
same time, eliminating one of the two fields involved. In the case of
\eqref{eq(1.4.1)}, we can eliminate the gauge parameter $\xi_{b_1b_2}$
by taking an exterior derivative of $E_{a,b_1b_2}$ which produces
\begin{equation}
    \partial_{[a_1}E_{a_2],b_1b_2} \equiv \partial_{[a_1}\omega_{a_2],b_1b_2}+\tfrac{1}{2}\,\varepsilon_{b_1b_2}{}^{c_1c_2}\,\partial_{[a_1|}\overline{G}_{c_1,c_2|a_2]}= 0\;. 
    \label{eq(1.4.3)}
\end{equation}
By contracting $a_1$ with $b_1$\,, we find that the term involving the dual
graviton vanishes due to the fact that we have anti-symmetrised derivatives
and $A_{ab}$ is symmetric. Thus, we find that
\begin{equation}
    \partial_{[a_1}\omega_{a_2],b_1b_2}\eta ^{a_1b_1}= 0\;,
    \label{eq(1.4.4)}
\end{equation}
which is the equation of motion for linearised gravity.

We can also write \eqref{eq(1.4.3)} as
\begin{equation}
    \tfrac{1}{2}\,\varepsilon_{b_1b_2}{}_{c_1c_2}\,\partial_{[a_1}\omega_{a_2],}{}^{b_1b_2}-\partial_{[a_1}\overline{G}_{[c_1,c_2]a_2]}= 0\;.
    \label{eq(1.4.5)}
\end{equation}
The first term is
$-\varepsilon_{b_1b_2}{}_{c_1c_2}\partial_{[a_1 }\partial^{b_1}h^{b_2 }{}_{a_2]}$
and so it vanishes when we contract it with $\eta^{a_1c_1}$\,. As a result, we find that
\begin{equation}
    \partial^{[a_1}\overline{G}_{[a_1,c_2]}{}^{a_2]}= 0\;,
    \label{eq(1.4.6)}
\end{equation}
which we recognise as the equation of motion for the dual graviton at the
linearised level, which agrees with the results of \cite{Glennon:2020qpt} where
the full non-linear equation of motion was found and its linearised version
was also given.

We will now carry out the same procedure for the duality relation involving
the dual graviton and the first higher dual graviton \eqref{eq(1.1.9)}.
Using the gauge transformations in \eqref{eq(1.2.10)} and \eqref{eq(1.2.11)}, we find that the duality relation becomes
\begin{equation}
    \overline{E}_{a , b_1 b_2} \equiv  \overline{G}_{a,b_1 b_2} + \varepsilon_a{}^{c_1 c_2 c_3}\,{{\overline{\overline {G}}}}_{c_1 , c_2 c_3 , b_1 b_2} -2\,\partial_a\partial_{(b_1}{\overline \xi}_{b_2)}-3\,e_1\, \varepsilon_a{}^{e_1e_2e_3}\,\partial_{e_1} \partial_{(b_1|} \Lambda_{e_2e_3, | b_2)}=0\;.
    \label{eq(1.4.7)}
\end{equation}
By taking two derivatives, we find that the gauge parameters disappear. We obtain
\begin{equation}
    \partial^{[c_1}\partial_{[b_1|}{\overline E}_{a,|b_2]}{}^{c_2]}
    \equiv \partial^{[c_1}\partial_{[b_1|}\overline{G}_{a,|b_2]}{}^{c_2]} + \varepsilon_a{}^{d_1 d_2 d_3}\,\partial^{[c_1}\partial_{[b_1}{{\overline{\overline {G}}}}_{[d_1 , d_2 d_3 ], b_2]}{}^{c_2]}   = 0\;,
    \label{eq(1.4.8)}
\end{equation}
which can also be written as
\begin{equation}
    \tfrac{1}{3!}\,\varepsilon_{e_1e_2e_3}{}^{a}\partial^{[c_1} \partial_{[b_1 |} \overline{G}_{a, | b_2]}{}^{c_2]}+
    \partial^{[c_1} \partial_{[b_1} {{\overline{\overline {G}}}}_{[e_1 , e_2e_3], b_2]}{}^{c_2]}=0\;.
    \label{eq(1.4.9)}
\end{equation}
The first term vanishes if we sum over $e_1$ and $b_1$ and also $e_2$ and $b_2$\,. 
From this, we obtain
\begin{equation}
    \partial^{[c_1} \partial_{[b_1} {{\overline{\overline {G}}}}^{[ b_1 , b_2 e ]}{}_{, b_2]}{}^{c_2]}=0\;,
    \label{eq(1.4.10)}
\end{equation}
which is indeed the correct equation of motion for the first higher 
dual graviton in four spacetime dimensions \cite{Bekaert:2002dt}. 

Clearly, the duality equation \eqref{eq(1.1.11)} between the graviton and the
first higher dual graviton will also lead to the same equations since it can be
deduced from the above duality relations. However, it is instructive to treat
this in the same way. Using the gauge transformation \eqref{eq(1.2.11)}, we find
that this duality relation is given by
\begin{equation}
    E'\,^{a,}{}_{b_1b_2} \equiv \omega^{a,}{}_{b_1b_2}+3\,\overline{\overline {G}}{}_{[b_1,b_2 c],}{}^{ca} -2\,\partial^a \partial_{[b_1}\xi_{b_2]}-3\,e_2\,\partial^{(a}   \partial_{[b_1}\Lambda_{b_2 c],}{}^{c)}=0\;.
    \label{eq(1.4.11)}
\end{equation}
The gauge parameter $\xi^a$ is then eliminated by taking a derivative as follows:
\begin{equation}
    \partial_{[a_1}E'_{a_2],}{}^{b_1b_2} \equiv \partial_{[a_1}\omega_{a_2],}{}^{b_1b_2}-3\,\partial_{[a_1|}\overline{\overline {G}}{}^{[b_1,b_2 c],}{}_c{}_{|a_2]}  - \tfrac{3}{2}\,e_2\,\partial_{c}\, \partial_{[a_1}\partial^{[b_1}\Lambda^{b_2 c],}{}_{a_2]}=0\;.
    \label{eq(1.4.12)}
\end{equation}
Contracting $a_1$ and $b_1$ allows us to discard the first term as it is
the equation of motion for linearised gravity. We are left with the equation
\begin{equation}
    3\,\partial^{[c|}\,\overline{\overline {G}}{}_{[c,bd],}{}^d{}^{|a]} - \tfrac{3}{2}\,e_2\,\partial^{d}\partial^{[c}\partial_{[c}\Lambda_{b  d] , }{}^{a]}=0\;.
    \label{eq(1.4.13)}
\end{equation}
Then, taking one more derivative, we can eliminate the last gauge parameter to arrive at
\begin{equation}
    \partial^{[a_1} \partial^{[c}\, \overline{\overline {G}}_{[c, bd] ,}{}^{d]a_2 ]}=0\;,
    \label{eq(1.4.14)}
\end{equation}
This is the correct equation of motion for the first higher dual graviton at the 
linearised level that we have also found in \eqref{eq(1.4.10)}.

The equations of motion \eqref{eq(1.4.4)}, \eqref{eq(1.4.6)} and
\eqref{eq(1.4.10)} for the graviton, the dual graviton, and the first 
higher dual graviton, with their respective symmetry 
types $\ytableaushort{\null\null}$\;, $\ytableaushort{\null\null}$\;, and
$\ytableaushort{\null\null\null,\null}$\;, are tracelessness equations that
may be written in the form
\begin{align}
    \mathrm{Tr}_{12}K_{a_1a_2,b_1b_2}=0\;,\qquad\mathrm{Tr}_{12}\overline{K}_{a_1a_2,b_1b_2}=0\;,\qquad\mathrm{Tr}_{12}^2\overline{\overline{K}}_{a_1a_2a_3,b_1b_2,c_1c_2}=0\;,
\end{align}
where $\mathrm{Tr}_{ij}$ denotes a trace over columns $i$ and
$j$ in a given Young diagram, and where we have introduced the curvature
tensors for each field. They are given explicitly by
\begin{gather}
    K_{a_1a_2,b_1b_2}\equiv\partial_{[a_1}\omega_{a_2],b_1b_2}\;,\qquad\overline{K}_{a_1a_2,b_1b_2}\equiv\partial_{[a_1}\overline{G}_{[b_1,b_2]a_2]}\;,\\
    \overline{\overline{K}}_{a_1a_2a_3,b_1b_2,c_1c_2}\equiv\partial_{[c_1}\partial_{[b_1}\overline{\overline{G}}_{[a_1,a_2a_3],b_2]c_2]}\;.
\end{gather}

As we have explained above, the duality relations only hold modulo gauge 
transformations, although the equations of motion derived from them hold exactly.
One of the points of this paper is to obtain the extra fields that are required to
have duality relations that also hold exactly. We will find evidence that these
extra fields are contained in the second fundamental representation of $A_1^{+++}$,
denoted $\ell_2$\,. The content of this representation can be deduced by enlarging
the $A_1^{+++}$ algebra by attaching an additional node to the node labelled 2 in
the $A_1^{+++}$ Dynkin diagram, and then by taking only the generators of this enlarged
algebra that have level one with respect to this new node. One may then deduce the
commutation relations between generators in the adjoint and $\ell_2$ representations of
$A_1^{+++}$ by using the fact that the level is preserved and that the Jacobi identities
must hold. One can then add new fields corresponding to the $\ell_2$ generators and 
deduce their $A_1^{+++}$ transformations from their commutation relations. As the role
of the new fields is to soak up the gauge transformations in the duality relations, 
the next step must be to propose their gauge transformations. 
This involves writing down the variation of
the $\ell_2$ fields in terms of the derivative, which belongs to $\ell_1$\,, acting
on the gauge parameters that also belong to the $\ell_1$ representation.
This transformation can be deduced using level matching and group theory.
Given these transformations, one can then finally obtain new duality relations which
hold as exact equations, at least in principle, and in detail at low levels. We leave 
this calculation to a future paper.

\section{Higher dualisations of linearised gravity}
\label{section:off-shell}

In this section we give an action principle in four dimensions for the first
higher dual graviton $A_{ab,cd}$ whose equations of motion and gauge
transformations were obtained from the non-linear realisation of
{$A_1^{+++}\ltimes\ell_1$} in the previous section.
We will only be concerned with free dynamics and we will build the action
principle for the dual field $A_{ab,cd}\equiv A_{[ab],cd}\equiv A_{ab,(cd)}$ 
using the off-shell dualisation procedure proposed in \cite{Boulanger:2012df}.
In that paper, a field-theoretical interpretation was given for an infinite
subset of $E_{11}$ generators that transform in the $\mathrm{GL}(11)$-irreducible
representations whose Young tableaux are given in column notation as
\begin{equation}
    \big\{\mathbb{Y}[8,1]\,,\mathbb{Y}[9,8,1]\,,\mathbb{Y}[9,9,8,1]\,, \ldots\big\}
\end{equation}
where the $\mathrm{GL}(11)$ Young diagram $\mathbb{Y}[8,1]$ of the dual graviton
may have an unbounded number of columns of height nine glued to the left of it. 
It was argued in \cite{Boulanger:2012df} that gauge fields transforming in
these $\mathrm{GL}(11)$ representations enter higher and higher dual off-shell
formulations of linearised gravity. This will now be made quantitative by
working at the first few levels of dualisation with precise action principles.

In what follows, we first recall the basic ideas behind the parent action
procedure to derive dual actions for linearised gravity in any dimension $D$\,,
and then we will direct our attention to the four-dimensional case for which
$A_1^{+++}$ is the relevant Kac-Moody algebra.

\subsection{From the graviton to the dual graviton}\label{section:grav2dualgrav}

Off-shell dualisation of linearised gravity $h_a{}^b$ around 
$D$-dimensional Minkowski space-time was initiated in 
\cite{West:2001as} and \cite{West:2002jj}.
This was investigated further in \cite{Boulanger:2003vs} where the
authors made contact with the Curtright action \cite{Curtright:1980yk} 
and generalised this duality to higher-spin fields with spin $s>2$\,. 
Although the analysis of \cite{West:2001as} began with the fully
non-linear Einstein-Hilbert action, it is only for its linearisation 
that one can make the dual graviton and all of its 
higher dual generalisations appear off-shell \cite{Boulanger:2012df}. 
Following the original idea \cite{West:2001as}, consider the second order
Einstein-Hilbert action based on the vielbein $e_{\mu}{}^{a}$\,:
\begin{align}
\label{EH}
  S_{\rm EH}[e_{\mu}{}^{a}] \ = \ -
  \int \mathrm{d}^D x\;e\left[\Omega^{ab,c}(e)\,\Omega_{ab,c}(e)+
  2\,\Omega^{ab,c}(e)\,\Omega_{ac,b}(e) - 4\, \Omega_{ab,}{}^b(e)\,\Omega^{ac,}{}_{c}(e)\right]\;,  
 \end{align}
where $e := {\rm det}(e_\mu{}^a)$ and $\Omega_{ab,}{}^{c}(e) := 
 2\,e_{a}{}^{\mu}\,e_{b}{}^{\nu}\,\partial_{[\mu}e_{\nu]}{}^{c}$\;.
This form of the Einstein-Hilbert action can be recast into first-order form by
introducing an auxiliary field $Y_{ab;c}=Y_{[ab];c}\,$ and then by considering
the parent action \cite{West:2001as}
\begin{align}
\label{first}
  S[Y_{ab;c}\,,e_{\mu}{}^{a}] \ = \ -2\,\int \mathrm{d}^Dx \; e
  \left[\Omega_{ab,c}(e)\,Y^{ab;c}-\tfrac{1}{2}\,Y_{ab;c}\,Y^{ac;b}
  +\tfrac{1}{2(D-2)}\,Y_{ab;}{}^{b}\,Y^{ac;}{}_{c}\right]\;. 
\end{align}
Indeed, the field equation of $Y_{ab;c}$ can solved for $Y_{ab;c}$
in terms of $\Omega(e)$ which yields
\begin{align}
\label{Ysol}
  Y_{ab;c}(e) \ = \
  \Omega_{ab,c}-2\,\Omega_{c[a,b]}+4\,\eta_{c[a}\Omega_{b]d,}{}^{d}\;.  
\end{align}
After inserting (\ref{Ysol}) into (\ref{first}), one
recovers the Einstein-Hilbert action (\ref{EH}). In
fact, the action in (\ref{first}) coincides with the standard first
order action for gravity where the spin connection is an independent field, 
up to a field redefinition which replaces the spin connection by the
$Y_{ab;c}$ field. The parent action in (\ref{first}) is
manifestly invariant under diffeomorphisms and local Lorentz transformations. 
In terms of the Hodge dual field 
\begin{align}
\label{dualY}
  Y_{c_1\ldots c_{D-2};}{}^d \; := \; 
  -\tfrac{1}{2}\,\varepsilon_{abc_1\cdots c_{D-2}}\,
  Y^{ab;d}\;, 
\end{align}
the parent action linearised around Minkowski spacetime, 
where $e_{\mu}{}^{a} = \delta_{\mu}{}^{a}+ h_{\mu}{}^{a}\,$,
reads 
\begin{align}
    \label{firstdual1}
    \!S[Y_{a_1\ldots a_{D-2};}{}^b\,,\,h_{ab}] =  -\tfrac{2}{(D-2)!}&\int\!
    \mathrm{d}^Dx\,\Big[\varepsilon^{abc_1\ldots
    c_{D-2}}\,Y_{c_1\ldots c_{D-2};}{}^{c}\,\Omega_{ab,c}(h)+\tfrac{D-3}{2(D-2)}\;
    Y^{c_1\ldots c_{D-2};b}\,Y_{c_1\ldots c_{D-2};b}\nonumber\\
    &-\tfrac{D-2}{2}\;Y^{c_1\ldots c_{D-3}a;}{}_{a}\,Y_{c_1\ldots c_{D-3}b;}{}^{b}
    +\tfrac{1}{2}\;Y^{c_1\ldots c_{D-3}a;b}\, Y_{c_1\ldots c_{D-3}b;a}\Big]\;,
\end{align}
where $\Omega_{ab,c}(h):=2\,\partial_{[a}h_{b]c}\,$ and the field
$h_{ab}$ has no symmetry on its two indices.
The equation of motion for $h_{ab}$ yields
\begin{align}
\label{inte}
  \partial_{[a_1}Y_{a_2\ldots a_{D-1}];b} \, = \, 0\;.
\end{align}
The Poincar\'e lemma implies that the dual field
$Y_{a_1\ldots\,a_{D-2};b}$ is the curl of a
potential $C_{a_1\ldots\,a_{D-3};b}$\,. This new field is
completely antisymmetric in its first $D-3$ indices but it has 
no definite $\mathrm{GL}(D)$ symmetry otherwise:
\begin{align}
\label{intesol}
  Y_{a_1\ldots a_{D-2};b} \ = \
  \partial_{[a_1}C_{a_2\ldots a_{D-2}];b}\;.  
\end{align}
Inserting this back into the linearisation of \eqref{firstdual1} produces
a consistent quadratic action $S[C]\,$ that describes linearised gravity by
construction. Note that the field $h_{ab}$ acted as a Lagrange multiplier for
the constraint \eqref{inte}.
It is not an auxiliary field like $Y_{a_1\ldots a_{D-2};b}$ is,
but the dual action obtained by substituting \eqref{intesol}
inside the parent action \eqref{firstdual1} is classically equivalent to the 
original linearised Einstein-Hilbert action. The reader might want to see
\cite{Fradkin:1984ai} for more comments on this issue.

Until now, the dual field $C_{a_1\ldots a_{D-3};b}$ as defined in
(\ref{intesol}) does not transform in any irreducible $\mathrm{GL}(D)$
representation since $Y_{a_1\ldots a_{D-2};b}$ 
does not have any irreducible $\mathrm{GL}(D)$ symmetry property.
However, one may check \cite{West:2002jj,Boulanger:2003vs} that, 
after inserting (\ref{intesol}) into (\ref{firstdual1}), 
the resulting action $S[C]$ is invariant under a shift symmetry 
inherited from the local Lorentz symmetry
\begin{align}
\label{stuckel}
  \delta_{\Lambda}C_{a_1\ldots a_{D-3};b} \ = \
  -\Lambda_{a_1\ldots a_{D-3}b}\;,  
\end{align}
with a completely antisymmetric $(D-2)$-form gauge parameter 
that is nothing but the Hodge dual of the local Lorentz 
parameter $\Lambda_{ab}\,$. 
In particular, in $D=11$ dimensions, the 9-form component
of the field $C_{a_1\ldots a_8;b}$ drops out from the action due 
to the above gauge symmetry. This gives rise to a dual action in
terms of the other component of $C_{a_1\ldots a_8;b}$ denoted by
$A_{a_1\ldots a_8,b}$ that we call the dual graviton 
\cite{West:2001as,West:2002jj,Boulanger:2003vs}. 
In the antisymmetric convention for Young tableaux, the $\mathrm{GL}(11)$
irreducibility condition of the dual graviton is the
over-antisymmetrisation identity
\begin{align}
    A_{[a_1\ldots a_8,b]} \equiv 0\;.
\end{align}
To summarise, the dual graviton field 
$A_{a_1\ldots a_{D-3},b}$ in $D$ dimensions is antisymmetric in
its first $D-3$ indices and it obeys the irreducibility constraint
$A_{[a_1\ldots a_{D-3},b]} \equiv 0\,$. The dual graviton is 
a $\mathrm{GL}(D)$-irreducible tensor of type $\mathbb{Y}[D-3,1]\,$. 

It is important to stress the fact that the dynamics of 
linearised gravity around Minkowski space-time, as given by the
variational principle based on the original Fierz-Pauli action, 
can equivalently be described from the dual action principle 
$S[A_{a_1\ldots a_{D-3},b}]$ given in \cite{Boulanger:2003vs}. 
The reason is that both the Fierz-Pauli action and the dual action 
appear upon elimination of different fields from the same parent action. 
Moreover, as explained in \cite{Boulanger:2003vs}, the dual graviton
in four dimensions is a symmetric field $A_{ab}=A_{(ab)}$ and the
dual action $S[A_{ab}]$ reproduces the standard Fierz-Pauli action.
In $D=4$, one concludes that 
``Fierz-Pauli is dual to Fierz-Pauli'' \cite{Boulanger:2003vs}.

In the next part, we review the dualisation procedure first explained  
in \cite{Boulanger:2012df}, which takes the dual action
$S[A_{a_1\ldots a_{D-3},b}]$ and produces a dual action featuring the first 
higher dual graviton $A_{a_1\ldots a_{D-2},b_1\ldots b_{D-3},c}$ 
as well as an extra field that cannot be eliminated from the action.
In four dimensions, the first higher dual 
graviton $A_{a_1a_2,bc}$ corresponds to the $A_1^{+++}$ generator 
$R^{a_1a_2,(bc)}$ at level 2. Therefore, this approach makes direct
contact with the previous section where the non-linear realisation of 
{$A_1^{+++}\ltimes\ell_1$} was reviewed.
The extra fields that enter each higher dual action principle will then be
shown to be closely correlated with the $\ell_2$ representation of
$A_1^{+++}$\,.
Although they are not needed in order to write down self-duality 
equations, they are necessary for the off-shell formulation of various
generations of higher dual graviton fields.

\subsection{The first higher dual graviton in four dimensions}
\label{subsection:firsthigher}

\paragraph{Action principle.}
As explained in \cite{Boulanger:2003vs}, around 
Minkowski spacetime of dimension $D=4$, the dual graviton
$A_{a_1\ldots a_{D-3},b}$ is a symmetric 
rank-2 tensor $A_{ab}\equiv A_{(ab)}$ and the dual 
action is just the Fierz-Pauli action given as 
follows, up to boundary terms that we neglect:
\begin{align}
    S_{\mathrm{FP}}[A_{ab}] = \int \mathrm{d}^4x \:
    \Big[-\tfrac{1}{2}\,\partial_a A_{bc}\,\partial^a A^{bc}\,
    +\tfrac{1}{2}\,\partial_a A_b{}^b\,\partial^a A_c{}^c\,
    - \partial_a A^{ab}\,\partial_b A + \partial_a A^{ab}\,\partial^c A_{cb} \Big]\;. 
    \label{originPF}
\end{align}
We stress that the curl $\Omega_{ab,c}(A):=2\,\partial_{[a}A_{b]c}$ is \emph{not} featured in this formulation of the Fierz-Pauli action. 
Instead, it features the full gradient $G_{a;bc}(A):=\partial_{a}A_{bc}\,$ 
without any antisymmetrisation over indices.
As proposed in \cite{Boulanger:2012df}, we define the following parent action $S[G_{a;bc}\,,D_{ab;}{}^{cd}]$\,:
\begin{align}
    S= \int \! \mathrm{d}^4x
    \Big[-\tfrac{1}{2}\,G_{a;bc}\,G^{a;bc}
    +\tfrac{1}{2}\,G_{a;c}{}^c \,G^{a;b}{}_b
    -G_{a;}{}^{ab}\,G_{b;c}{}^c
    +G_{a;}{}^{ab}\,G^{c;}{}_{cb}
    +G_{a;bc}\,\partial_{d}D^{da;bc}\Big]
    \label{parent}
\end{align}
featuring the two independent fields $G_{a;bc}=G_{a;(bc)}$ and 
$D_{ab;}{}^{cd}=D_{[ab];}{}^{cd}=D_{ab;}{}^{(cd)}$\,.
The latter of these two fields is defined up to a gauge transformation
\begin{align}
    \delta_\Theta D_{ab;}{}^{cd} = \partial^e\Theta_{eab;}{}^{cd}\;,
    \quad
    \Theta_{eab;}{}^{cd}\equiv\Theta_{[eab];}{}^{cd}\equiv\Theta_{eab;}{}^{(cd)}\;,
    \label{deltaDTheta}
\end{align}
which preserves the parent action. In fact, 
since the original Fierz-Pauli action \eqref{originPF} is invariant under 
the gauge transformation
\begin{align}
    \delta A_{ab} = 2\,\partial_{(a}\epsilon_{b)}\;,
\end{align}
it is easy to see that the parent action \eqref{parent}
is invariant under the combined transformations
\begin{align}
    \delta G_{a;bc} &= 2\,\partial_a\partial_{(b}\epsilon_{c)}\;,
    \\
    \delta D_{ab;}{}^{cd}&= \partial^e\Theta_{eab;}{}^{cd} + 2\, \eta^{cd} \pd_{[a} \epsilon_{b]} + 4\, \delta_{[a}{}^{(c} \pd_{b]} \epsilon^{d)}\;.
    \label{invaD}
\end{align}

On the one hand, one can vary the parent action \eqref{parent} with respect 
to the $\mathrm{GL}(4)$-reducible field $D^{da;bc}$ that acts as a Lagrange multiplier
for the constraint $\partial_{[a}G_{d];bc}=0$\,. 
This constraint is identically solved by 
\begin{align}
    G_{a;bc}=G_{a;bc}(A):=\partial_{a}A_{bc}\;,
\end{align}
for some symmetric tensor $A_{bc}\,$.
Substituting $G_{a;bc}(A)$ for $G_{a;bc}$ inside the parent action
\eqref{parent} reproduces the original Fierz-Pauli action \eqref{originPF}.

On the other hand, in the parent action \eqref{parent}, 
the independent field $G_{a;bc}$ can be considered to be an 
auxiliary field. Its equation of motion
\begin{align}
    0=\partial^{e}D_{ea;bc}-G_{a;bc}+\eta_{bc}\,G_{a;e}{}^e-\eta_{a(b}\,G_{c);e}{}^e
    -\eta_{bc}\,G_{e;a}{}^e+2\,\eta_{a(b}\,G^{e;}{}_{c)e}
\end{align}
can be solved algebraically to express $G_{a;bc}$ 
in terms of $D_{ab;}{}^{cd}$ as follows:
\begin{align}
    \frac{\delta S[G,D]}{\delta G^{a;bc}}\,=\,0 \quad \Longrightarrow\quad
    G_{a;bc} = \partial^{e}D_{ea;bc}+\tfrac{1}{2}\,\eta_{bc}\,\partial^{d}D_{ad;c}{}^c 
    +\tfrac{2}{3}\,\partial^{d}D_{ed;}{}^{e}{}_{(b}\;\eta_{c)a}\;.
\end{align}
Upon substituting this expression for $G_{a;bc}$ into the parent action 
\eqref{parent}, we obtain the following alternative description 
of linearised gravity around four-dimensional Minkowski spacetime:
\begin{align}
    S[D_{ab;}{}^{cd}] = \int \mathrm{d}^4x \;
    \Big[ \tfrac{1}{2}\,\partial^{a}D_{ab;}{}^{cd}\,\partial_{e}D^{eb;}{}_{cd}
    -\tfrac{1}{3}\,\partial^{a}D_{ea;}{}^{eb}\,\partial_{c}D^{dc;}{}_{db}
    +\tfrac{1}{4}\,\partial^{a}D_{ab;}{}^{c}{}_c\,\partial_{e}D^{be;}{}_{d}{}^d
    \Big]\;.
    \label{firsthigher}
\end{align}
This action is invariant under the gauge transformation \eqref{invaD}.
We emphasise that \eqref{firsthigher} describes the same free
graviton dynamics as the Fierz-Pauli action \eqref{originPF}. 
The reason is that both action principles arise from the same parent 
action $S[G,D]$ when it is extremised with respect to one field or the other.
Note that the spectrum of fields is in one-to-one correspondence with those
that are obtained by taking the tensor product of a 2-form with a symmetric 
rank-2 tensor. This is depicted in terms of Young tableaux as follows:
\begin{equation}\label{decompD}
    D_{ab;}{}^{cd}
    \quad\sim\quad
    \ytableaushort{a,b}\quad\otimes\quad\ytableaushort{cd}
\end{equation}

\paragraph{Relation to $A_1^{+++}$.}
In what follows, we decompose the dual field 
$D_{ab;}{}^{cd}$ into $\mathrm{GL}(4)$-irreducible components to allow for direct
contact with the $A_1^{+++}$ algebra. 
In particular, we will show how the first higher dual graviton 
$A_{ab,cd}$ that transforms in the $\mathrm{GL}(4)$-irreducible representation
of type $\mathbb{Y}[2,1,1]$ associated with the $A_1^{+++}$ generator
$R^{\,ab,(cd)}$ is contained in the $\mathrm{GL}(4)$-reducible field
$D_{ab;}{}^{cd}$ in \eqref{decompD}.
This will differ from the first section where we were concerned with duality relations modulo certain gauge transformations. In this section, we find an
off-shell formulation of linearised gravity in terms of $A_{ab,cd}$ and an
extra field $\widehat{Z}^{abc,d}$ which are both contained inside the
$D_{ab;}{}^{cd}$ field given previously.

The $\mathrm{GL}(4)$-irreducible decomposition of $D_{ab;}{}^{cd}$ reads
\begin{align}
    D_{ab;}{}^{cd} = X_{ab;}{}^{cd} + 4\, \delta_{[a}{}^{(c}\,
    {Z}_{b];}{}^{d)}\;,
    \quad X_{ab;}{}^{cb}\equiv 0 \equiv {Z}_{a;}{}^{a}\;,
\end{align}
with inverse formulas
\begin{align}
    X_{ab;}{}^{cd} = D_{ab;}{}^{cd} + \delta_{[a}{}^{(c}D_{b]e;}{}^{d)e}\; , \quad {Z}_{a;}{}^b = - \tfrac{1}{4} D_{ac;}{}^{bc}\; .
\end{align}
In terms these fields, the dual gravity action 
\eqref{firsthigher} becomes
\begin{align}
    S[X_{ab;}{}^{cd},{Z}_{a;}{}^{c}]=\int & \mathrm{d}^4x\,\Big[ 
    \tfrac{1}{2}\,\partial_a X^{ab;}{}_{cd}\,\partial^e X_{eb;}{}^{cd}
    -\tfrac{1}{4}\,\partial^a X_{ab;}{}^c{}_c\,\partial_e X^{eb;}{}_d{}^d
    \nonumber \\
    & +2\,\partial_a {Z}^{a;b}\partial_c {Z}_{b;}{}^c
    -\tfrac{10}{3}\,\partial_a \,{Z}^{a;b}\partial^c {Z}_{c;b} 
    +\partial_c {Z}_{a;b}\,\partial^c {Z}^{a;b}
    \nonumber\\
    &+\,2\,\partial_b {Z}^{[a;b]}\,
    \partial^c X_{ac;}{}^d{}_d 
    - 2\,\partial^c {Z}^{a;b} \, \partial^e X_{ae;bc}
    \Big]\;.
    \label{actionXZ}
\end{align}
Hodge dualising $X_{ab;}{}^{cd}$ and $Z_{a;}{}^{c}$ on their lower 
indices produces the $\mathrm{GL}(4)$-irreducible fields
\begin{equation}
    A^{ab,cd} := -\tfrac{1}{2}\,\varepsilon^{abij}\,X_{ij;}{}^{cd}\;,\qquad
    \widehat{Z}^{abc,d} := \varepsilon^{abce}\,Z_{e;}{}^d\;,
    \label{level1XA}
\end{equation}
with inverse relations
\begin{equation}
    X_{ab;}{}^{cd}=\tfrac{1}{2}\,\varepsilon_{abij}\,A^{ij,cd}\;,
    \qquad
    Z_{a;}{}^{e}=\tfrac{1}{6}\,\varepsilon_{abcd}\,\widehat{Z}^{bcd,e}\;.
    \label{level1XA-inverse}
\end{equation}
These fields satisfy $\mathrm{GL}(4)$ irreducibility conditions in the antisymmetric
convention for Young tableaux. That is to say, they satisfy the 
over-antisymmetrisation identities:
\begin{gather}
    A^{[ab,c]d} \equiv 0\;, \qquad\qquad\qquad  \widehat{Z}^{[abc,d]}\equiv 0\;,
\end{gather}
where $A^{ab,cd} = A^{[ab],cd}=A^{ab,(cd)}$
and $\widehat{Z}^{abc,d} = \widehat{Z}^{[abc],d}$\,.
The reader will recognise that the field $A^{ab,cd}$ possesses all 
the symmetries of the first higher dual graviton defined in the 
previous section. It corresponds to the generator $R^{ab,cd}$ of $A_1^{+++}$\,.
We also see that the field $\widehat{Z}^{abc,d}$ is required for the
action principle to exist. In terms of the two $\mathrm{GL}(4)$-irreducible fields $A^{ab,cd}$ and $\widehat{Z}^{abc,d}$\,, the dual gravity action 
\eqref{firsthigher} now reads
\begin{align}
    S[A^{ab,cd}&,\widehat{Z}^{abc,d}]=\int \mathrm{d}^4x\,\Big[ 
    -\tfrac{3}{4}\,\partial_e A_{ab,cd}\,\partial^{[e} A^{ab],cd}
    +\tfrac{3}{8}\,\partial_d A_{ab,c}{}^{c}\,\partial^{[d} A^{ab],e}{}_{e}
    \nonumber\\
    &
    \qquad\qquad\qquad\quad\quad
    -\tfrac{3}{2}
    \,\partial_d A_{ab,c}{}^{c}\,\partial^{[d}\widehat{Z}^{ab]e,}{}_{e}
    +\partial_e A_{ab,cd}\,\partial^d\widehat{Z}^{eab,c}
    \label{actionAZ}\\
    &
    + \partial_d\widehat{Z}_{abc,}{}^{c}\,\partial_e\widehat{Z}^{abd,e}
    -\tfrac{1}{3}\,\partial_e\widehat{Z}_{abc,d}\,\partial^d\widehat{Z}^{abc,e}
    -\tfrac{5}{3}\,\partial_e\widehat{Z}_{abc,d}\,\partial^c\widehat{Z}^{abe,d}
    +\tfrac{7}{18}\,\partial_e\widehat{Z}_{abc,d}\,\partial^e\widehat{Z}^{abc,d}
    \Big]\;.\nonumber
\end{align}
This action is invariant under the gauge transformations
\begin{align}
    \delta{}A^{ab,cd}
    &=
    -4\partial^{[a}\lambda^{b]cd}
    +\partial^{[a|}\mu^{cd,|b]}
    +2\partial^{(c}\mu^{d)[a,b]}
    -\varepsilon^{ij[a(c}\eta^{d)b]}\partial_i\epsilon_j
    -\tfrac{1}{2}\varepsilon^{ijab}\eta^{cd}\partial_i\epsilon_j\;,
    \\
    \delta\widehat{Z}^{abc,d}
    &=
    {3}\,\partial^{[a|}\mu^{d|b,c]}
    +\tfrac{1}{4}\,\varepsilon^{abce}\,\partial^d\epsilon_e
    +\tfrac{3}{4}\,\varepsilon^{abce}\,\partial_e\epsilon^d
    -\tfrac{1}{4}\,\varepsilon^{abcd}\,\partial^e\epsilon_e\;,
    \label{gaugevariahatZ}
\end{align}
where the gauge parameters $\lambda^{abc}$ and $\mu^{ab,c}$ are 
$\mathrm{GL}(4)$-irreducible:
\begin{align}
    \lambda^{abc}=\lambda^{(abc)}\quad \sim\quad 
    \ytableausetup{smalltableaux}\ytableaushort{abc}
    \;,\qquad 
    \mu^{ab,c} = \mu^{(ab),c} 
    \quad \sim\quad 
    \ytableausetup{smalltableaux}\ytableaushort{ab,c}
    \;,\quad \mu^{(ab,c)} \equiv 0\;.  
\end{align}
One may, of course, equivalently use the manifestly antisymmetric
convention for Young tableaux in expressing the mixed-symmetric gauge
parameter by taking
\begin{align}
    m^{ab,c}:=\,2\,\mu^{c[a,b]}
    \; \sim\; 
    \ytableausetup{smalltableaux}\ytableaushort{ac,b}
    \quad , \qquad m^{[ab,c]}\equiv 0
    \qquad \Leftrightarrow\qquad
    \mu^{ab,c} = -\tfrac{2}{3}\,m^{c(a,b)}\;.
\end{align}
In terms of this equivalent representation for the gauge parameter, 
one has
\begin{align}
    \delta{}A^{ab,cd}
    &=
    -4\partial^{[a}\lambda^{b]cd}
    -\tfrac{2}{3}\,\partial^{[a}m^{b](c,d)}
    +\partial^{(c|}m^{ab,|d)}
    -\varepsilon^{ij[a(c}\eta^{d)b]}\partial_i\epsilon_j
    -\tfrac{1}{2}\,\varepsilon^{ijab}\eta^{cd}\partial_i\epsilon_j\;,
    \label{gaugevariaAbis}
    \\
    \delta\widehat{Z}^{abc,d}
    &=
    \tfrac{3}{2}\,\partial^{[a} m^{bc],d}
    +\tfrac{1}{4}\,\varepsilon^{abce}\,\partial^d\epsilon_e
    +\tfrac{3}{4}\,\varepsilon^{abce}\,\partial_e\epsilon^d
    -\tfrac{1}{4}\,\varepsilon^{abcd}\,\partial^e\epsilon_e\;.
    \label{gaugevariahatZbis}
\end{align}
Notice that $\lambda^{abc}$ and $m^{ab,c}\sim\mu^{ab,c}$ match the gauge parameters
$\Lambda_{a_1a_2a_3}$ and $\Lambda_{a_1a_2,b}$ at level 2 in the $\ell_1$
representation of $A_1^{+++}$ given in \eqref{l1:gaugeparams}.
Up to trivial gauge parameter redefinitions, the transformation 
law of $A^{ab,cd}$ with respect to $m^{ab,c}$ and $\lambda^{abc}$
fully agrees with \eqref{eq(1.2.11)}.

Off-shell dualisation is a different approach to the $A_1^{+++}$ non-linear
realisation presented in the previous section. 
The gauge transformations found here contain extra
terms compared with \eqref{eq(1.2.11)} which are due to the extra field
$\widehat{Z}^{abc,d}$ in \eqref{actionAZ}.
We will soon observe that the $\ell_2$ representation of $A_1^{+++}$ is closely
related to extra fields that appear during off-shell dualisation. Therefore, we
expect to obtain \eqref{gaugevariaAbis} and \eqref{gaugevariahatZbis}
from the non-linear realisation by modifying it in a suitable way that 
incorporates the $\ell_2$ representation.

The gauge parameters $\lambda^{abc}$ and $\mu^{ab,c}$ arise from the decomposition 
\begin{align}
 \Theta_{abc;}{}^{de} = 2\,\varepsilon_{abci}\,(-\lambda^{dei} + \mu^{de,i}) \;,\label{thetadecomposition}  
\end{align}
so that the $\Theta_{abc;}{}^{de}$ part of the gauge transformation 
in \eqref{invaD} reads
\begin{align}
    \delta_{\Theta} D_{ab;}{}^{ef} = -2\,\varepsilon_{abcd}\,
    (\partial^c\lambda^{def} - \partial^c\mu^{ef,d})\;.
\end{align}

We stress that this dual action principle \eqref{actionAZ}--\eqref{gaugevariahatZ} 
describes equivalent dynamics to the Fierz-Pauli action principle.
Namely, it propagates a single graviton in four-dimensional Minkowski spacetime.
It is an alternative off-shell description of linearised gravity and 
we will further analyse this action principle in Section \ref{section:repackage}.
In that section, in order to make contact with the Labastida formulation 
for a gauge field with the symmetries of the first higher-dual graviton, 
we need to change convention for Young tableau. 
We refer to Appendix \ref{appendix-phipsi}
for this change of convention.

Note that the field content of the theory 
$\{A^{ab,cd}\,,\widehat{Z}^{abc,d}\}$ 
is in one-to-one correspondence with the set of Young diagrams 
obtained in the tensor product 
\begin{align}
    \ytableausetup{smalltableaux}
    \ytableaushort{\null,\null}\quad\otimes\quad \ytableaushort{\null\null}
    \quad=\quad
    &
    \ytableaushort{\null\null\null,\null}
    \quad\oplus \quad
    \ytableaushort{\null\null,\null,\null}\quad .
    \label{firstlevel}
\end{align}
This depicts the set of $\mathrm{GL}(4)$-irreducible tensors that are
contained in the reducible tensor
\begin{align}
    \widetilde{D}^{ab;cd}:=\tfrac{1}{2}\,\varepsilon^{abij}\,D_{ij;}{}^{cd}\;.
    \label{Dtilde}
\end{align}

In Section \ref{section:repackage}, 
we will build two gauge invariant curvature tensors that do not vanish 
on-shell. 
Anticipating this (more technical) result, 
the curvature for the gauge field $A_{ab,cd}$ starts like 
\begin{align}
    \widehat{G}_{a_1a_2a_3,b_2b_2,c_1c_2} = 
    \partial_{a_1}\partial_{b_1}\partial_{c_1}A_{a_2a_3,(b_2c_2)} +\ldots\;,
\end{align}
where ``$\ldots$" is used to denote terms that involve the field $\widehat{Z}_{abc,d}$
and where it is understood that indices with the same letters are antisymmetrised.
For example,
$\partial_{a_1}V_{a_2}\equiv\tfrac{1}{2}\,(\partial_{a_1}V_{a_2} - \partial_{a_2}V_{a_1})\,$.
Then, in that same section, 
we show that the field equations for $A_{ab,cd}$  
are equivalent to 
\begin{align}
    \widehat{G}^{\,abc,}{}_{ab,de} = 0\;,
    \qquad\qquad
    \widehat{G}_{a_1a_2a_3,bc,}{}^{b}{}_d=0\;.
    \label{eomA}
\end{align}
As demonstrated in \cite{Bekaert:2002dt,Bekaert:2003az},
this form of field equation is precisely what one should have 
for a mixed-symmetric gauge field $A_{ab,cd}$ that propagates 
non-trivially in four dimensional Minkowski spacetime.
It is of higher-derivative type for a gauge field with more than 
two columns in its Young tableau representation, but a partial 
gauge-fixing procedure was found in \cite{Bekaert:2003az} that brings
such higher-derivative field equations down to the two-derivative 
equations (for bosonic fields) postulated in 
\cite{Labastida:1986ft,Labastida:1987kw}.

The first field equation in \eqref{eomA} is precisely of the form we have
seen before in \eqref{eq(1.4.14)}, except that now, since we have an action
principle for the first higher dual graviton, all the gauge invariant
quantities involve both $A_{ab,cd}$ and $\widehat{Z}_{abc,d}$ which duly reflects
the fact that the gauge transformations
\eqref{gaugevariaAbis} and \eqref{gaugevariahatZbis} are both expressed in terms
of the parameters $m^{ab,c}$ and $\epsilon_a$\,.
The gauge transformations are entangled as is typical when performing higher
off-shell dualisations \cite{Boulanger:2020yib}.
Now, not only do we have the first higher dual graviton field $A_{ab,cd}$\,,
but also the extra field $\widehat{Z}_{abc,d}$ that is required for our dual action 
principle to exist. Together, this pair of fields describes a single propagating
graviton. The extra field $\widehat{Z}_{abc,d}$ is not in the adjoint
representation of $A_1^{+++}$ but we will later see that it belongs to
the $\ell_2$ representation of $A_1^{+++}$ at level 1 in the decomposition of 
$A_1^{+++}$ with respect to its $\mathrm{GL}(4)$ subalgebra
(see Table \ref{A1+++-l2-level0123}).

\subsection{Field theoretical analysis at higher levels}
\label{section:higherdualgrav}

After having discussed, in great detail, off-shell dualisation from the dual graviton
$A_{ab}=A_{(ab)}$ to the first higher dual graviton 
$A_{ab,cd}\,$, we may now proceed to the next step in the off-shell dualisation
procedure. Recall that the dualisation procedure at level one transformed 
our set of fields from a symmetric tensor $A_{ab}$ to the $\mathrm{GL}(4)$-reducible field
${D}_{ab;}{}^{cd}$ whose Hodge dual \eqref{Dtilde} in four dimensions,
$\widetilde{D}^{ab;cd}$\,, contains 
the $\mathrm{GL}(4)$-irreducible fields $A^{ab,cd}$ and $\widehat{Z}^{abc,d}$  
with symmetry types $\mathbb{Y}[2,1,1]$ and $\mathbb{Y}[3,1]$, respectively, with Young tableaux given in \eqref{firstlevel}.

In order to dualise the field 
$A^{ab,cd}\sim\ytableausetup{smalltableaux}\ytableaushort{\null\null\null,\null}$\;, 
we build a parent action $S[G^{e;ab,cd},\widehat{Z}^{abc,d}]$ from
$S[A^{ab,cd},\widehat{Z}^{abc,d}]$ by treating $G^{e;ab,cd}$ as an independent field
that will be equal to the gradient $G^{e;ab,cd}(A):=\partial^{e}A^{ab,cd}$ 
after varying the parent action with respect to the Lagrange multiplier field 
$D_{ab;}{}^{cd,ef}$ that implements the constraint 
$\partial^{[e}G^{f];ab,cd}=0$\,. Therefore, the parent action at the next level
of dualisation is given schematically by
\begin{equation}
    S[D_{ab;}{}^{cd,ef},G^{e;ab,cd},\widehat{Z}^{abc,d}]
    :=
    S[G^{e;ab,cd},\widehat{Z}^{abc,d}]+\int \mathrm{d}^4x \; 
    G^{b;}{}_{cd,ef}\,\partial^{a}D_{ab;}{}^{cd,ef}\;.\label{deltaDTheta2}
\end{equation}
Both $G^{e;ab,cd}$ and $D_{ab;}{}^{cd,ef}$ have the 
$\mathrm{GL}(4)$-irreducible symmetries of $A^{cd,ef}$ in their final four indices, 
and $D_{ab;}{}^{cd,ef}=D_{[ab];}{}^{cd,ef}\,$.

As with \eqref{deltaDTheta}, 
the Lagrange multiplier field $D_{ab;}{}^{cd,ef}$ is defined up to
\begin{align}
    \delta_\Theta D_{ab;}{}^{cd,ef} = 
    \partial^i\Theta_{iab;}{}^{cd,ef}\;,
    \qquad \Theta_{iab;}{}^{cd,ef} = 
    \Theta_{[iab];}{}^{cd,ef} \;,
\end{align}
where $\Theta_{iab;}{}^{cd,ef}$ also shares 
the $\mathrm{GL}(4)$-irreducible symmetries of $A^{cd,ef}$ in its final
four indices.

The equation of motion for $G^{e;ab,cd}$ can be solved algebraically 
for $G^{e;ab,cd}$ in terms of $\widehat{Z}^{abc,d}$ and $D_{ab;}{}^{cd,ef}\,$. 
Then, as before, this expression may be substituted back 
into the parent action $S[D_{ab;}{}^{cd,ef},G^{e;ab,cd},\widehat{Z}^{abc,d}]$ 
to produce a new dual action $S[D_{ab;}{}^{cd,ef},\widehat{Z}^{abc,d}]$ 
that we will not write explicitly here. 
The $\mathrm{GL}(4)$-irreducible field content of the new field 
$D_{ab;}{}^{cd,ef}$ can be read off from its Hodge dual 
$\widetilde{D}^{ab;cd,ef}=\tfrac{1}{2}\,\varepsilon^{abij}\,D_{ij;}{}^{cd,ef}$
and the decomposition of its Young diagram:
\begin{align}
    \ytableausetup{smalltableaux}
    \ytableaushort{\null,\null}
    \quad\otimes\quad \ytableaushort{\null\null\null,\null}
    \quad
    \sim \quad
    &
    \ytableaushort{\null\null\null\null,\null\null}
    \quad\oplus \quad
    \ytableaushort{\null\null\null\null,\null,\null}
    \quad\oplus \quad 
    \ytableaushort{\null\null\null,\null\null,\null}
    \quad\oplus \quad 
    \ytableaushort{\null\null\null,\null,\null,\null}\;,\label{secondlevelforA}\\
    \Leftrightarrow~\quad
    \mathbb{Y}[2] ~\otimes ~\mathbb{Y}[2,1,1]\quad \sim \quad& 
    \mathbb{Y}[2,2,1,1]
    ~\oplus ~\mathbb{Y}[3,1,1,1]
    ~\oplus ~\mathbb{Y}[3,2,1]
    ~\oplus ~\mathbb{Y}[4,1,1]\;,\label{antisym}\\
    \Leftrightarrow\quad
    \mathbb{Y}(1,1) ~\otimes ~\mathbb{Y}(3,1)\quad \sim \quad&
    \mathbb{Y}(4,2)
    ~\oplus ~\mathbb{Y}(4,1,1)
    ~\oplus ~\mathbb{Y}(3,2,1)
    ~\oplus ~\mathbb{Y}(3,1,1,1)\;,\label{sym}\\
    \Leftrightarrow\qquad\qquad\quad
    D \qquad\qquad \sim\quad & \quad A \hspace{2.0cm} \widehat{Y} \hspace{2.1cm} \widehat{Z} \hspace{2.2cm} \widehat{W} \;.  
\end{align}
This demonstrates how to label Young tableaux either by the heights of their
columns as in \eqref{antisym}, or by the lengths of their rows as in \eqref{sym}.

Before explicitly performing this decomposition, we switch the convention
for Young tableaux to that where the final four indices of $D_{ab;}{}^{cd,ef}$
in the antisymmetric convention will be traded for
$D_{ab;}{}^{cde,f}:= -\tfrac{3}{2}D_{ab;}{}^{c(d,ef)}$ in the manifestly 
symmetric convention\footnote{The reason is purely technical: it comes from
the fact that we use the Mathematica package xTras \cite{Nutma:2013zea}
of the suite of Mathematica packages xAct that is able to implement
tracelessness constraints more easily than mixed Young tableaux
irreducibility constraints. The manifestly symmetric convention for Young
tableaux is explained in Appendix \ref{appendix-phipsi}.}.
with the over-symmetrisation identity $D_{ab;}{}^{(cde,f)}\equiv0$\,. 
This decomposition into $\mathrm{GL}(4)$-irreducible components becomes
\begin{equation}
    D_{a_1a_2;}{}^{c_1c_2c_3,d}
    =X_{a_1a_2;}{}^{c_1c_2c_3,d}
    +\delta_{[a_1}{}^{\langle{c_1}}Y_{a_2];}{}^{c_2c_3d\rangle}
    +\delta_{[a_1}{}^{\langle{c_1}}Z_{a_2];}{}^{c_2c_3,d\rangle}
    +\delta_{[a_1}{}^{\langle{c_1}}\delta_{a_2]}{}^{c_2}W^{c_3d\rangle}\;,
\end{equation}
where $\langle\,\cdots\rangle$ denotes projection onto irreducible components. Indices 
$a_1a_2$ are antisymmetric and indices $c_1c_2c_3$ are symmetric. The 
$\mathrm{GL}(4)$ irreducibility conditions are
\begin{equation}
    X_{a_1a_2;}{}^{(c_1c_2c_3,d)}=0\;,
    \qquad
    Z_{a;}{}^{(c_1c_2,d)}=0\;,
\end{equation}
together with the tracelessness constraints
\begin{equation}
    0\equiv{X}_{a_1b;}{}^{c_1c_2b,d}
    \equiv{X}_{a_1b;}{}^{c_1c_2c_3,b}
    \equiv{Y}_{b;}{}^{c_1c_2b}
    \equiv{Z}_{b;}{}^{c_1b,d}
    \equiv{Z}_{b;}{}^{c_1c_2,b}\;.\label{level2traceless}
\end{equation}
Projecting onto the symmetry of the final four indices, we find
\begin{align}
    D_{a_1a_2;}{}^{c_1c_2c_3,d}
    =\;&X_{a_1a_2;}{}^{c_1c_2c_3,d}
    +\delta_{[a_1}{}^{(c_1}Y_{a_2];}{}^{c_2c_3)d}
    -\delta_{[a_1}{}^{d}Y_{a_2];}{}^{c_1c_2c_3}\nonumber\\
    &
    +\delta_{[a_1}{}^{(c_1}Z_{a_2];}{}^{c_2c_3),d}
    +\delta_{[a_1}{}^{d}\delta_{a_2]}{}^{(c_1}W^{c_2c_3)}\;,\label{level2projection}
\end{align}
with inverse formulas
\begin{align}
    X_{a_1a_2;}{}^{c_1c_2c_3,d}
    =\;&D_{a_1a_2;}{}^{c_1c_2c_3,d}
    +\delta_{[a_1}{}^{d}D_{a_2]i;}{}^{c_1c_2c_3,i}
    -\tfrac{3}{5}\delta_{[a_1}{}^{(c_1}D_{a_2]i;}{}^{c_2c_3)d,i}\nonumber\\
    &+\tfrac{6}{5}\delta_{[a_1}{}^{(c_1}D_{a_2]i;}{}^{c_2c_3)i,d}
    -\tfrac{3}{5}\delta_{[a_1}{}^{d}\delta_{a_2]}{}^{(c_1}D_{ij;}{}^{c_2c_3)i,j}\;,\label{level2Xinverse}\\
    Y_{a;}{}^{c_1c_2c_3}
    =\;&D_{ai;}{}^{c_1c_2c_3,i}
    -\tfrac{1}{2}\delta_{a}{}^{(c_1}D_{ij;}{}^{c_2c_3)i,j}\;,\label{level2Yinverse}\\
    Z_{a;}{}^{c_1c_2,d}
    =\;&-\tfrac{6}{5}D_{ai;}{}^{c_1c_2i,d}
    -\tfrac{2}{5}D_{ai;}{}^{c_1c_2d,i}
    +\tfrac{4}{15}\delta_{a}{}^{(c_i}D_{ij;}{}^{c_2)di,j}
    -\tfrac{4}{15}\delta_{a}{}^{d}D_{ai;}{}^{c_1c_2i,j}\;,\label{level2Zinverse}\\
    W^{c_1c_2}
    =\;&-\tfrac{1}{3}D_{ij;}{}^{c_1c_2i,j}\;.\label{level2Winverse}
\end{align}
Previously, in order to dualise $A_{ab}$ off-shell, we decomposed
$D_{ab;}{}^{cd}$ into traceless components $\{X,Z\}$\,,
whose Hodge duals are the irreducible components of the Hodge dual 
$\widetilde{D}^{ij;cd}$ of $D_{ab;}{}^{cd}$\,. In order to make contact with
the E\,-theory literature expressed using fields and generators in
the antisymmetric convention, we will do something similar to $\{X,Y,Z,W\}$\,.
Hodge dualising all of them on their first blocks of indices creates
$\mathrm{GL}(4)$-irreducible fields in the symmetric convention, which may
then be written in the antisymmetric convention with fields labelled
$\{A\,,\widehat{Y},\widehat{Z},\widehat{W}\}$\,. The full calculation is given
in Appendix \ref{appendix-phipsi}, from which we find
\begin{align}
    X_{a_1a_2;}{}^{c_1c_2c_3,d} &:= -\tfrac{6}{5}\,\varepsilon_{a_1a_2b_1b_2}A^{(b_1|(b_2,d)|c_1,c_2,c_3)}\label{X2irred}\;,\\
    Y_{a;}{}^{c_1c_2c_3} &:= \varepsilon_{a_1b_1b_2b_3}\widehat{Y}^{b_2b_3(b_1,c_1,c_2,c_3)}\label{Y2irred}\;,\\
    Z_{a;}{}^{c_1c_2,d} &:= \tfrac{8}{5}\,\varepsilon_{ab_1b_2b_3}\widehat{Z}^{(b_1|b_3(b_2,d)|c_1,c_2)}\label{Z2irred}\;,\\
    W^{c_1c_2} &:= \varepsilon_{b_1b_2b_3b_4}\widehat{W}^{b_4b_3b_2(b_1,c_1,c_2)}\label{W2irred}\;.
\end{align}
with inverse relations
\begin{align}
    A^{a_1a_2,b_1b_2,c_1,c_2} :={}& -\tfrac{1}{10}\,\varepsilon^{ija_1a_2}X_{ij;}{}^{c_1c_2[b_1,b_2]}+\tfrac{1}{10}\,\varepsilon^{ij[a_1[b_1|}X_{ij;}{}^{c_1c_2|b_2],a_2]}\nonumber\\
    &{}-\tfrac{1}{5}\,\varepsilon^{ij[a_1(c_1}X_{ij;}{}^{c_2)a_2][b_1,b_2]} + \left(a_1a_2\leftrightarrow{}b_1b_2\right)\label{X2irredinverse}\;,\\
    \widehat{Y}^{a_1a_2a_3,c_1,c_2,c_3} :={}& -\tfrac{1}{6}\,\varepsilon^{ia_1a_2a_3}Y_{i;}{}^{c_1c_2c_3}-\tfrac{1}{2}\,\varepsilon^{i[a_1a_2(c_1}Y_{i;}{}^{c_2c_3)a_3]}\label{Y2irredinverse}\;,\\
    \widehat{Z}^{a_1a_2a_3,b_1b_2,c} :={}& -\tfrac{1}{15}\,\varepsilon^{ia_1a_2a_3}Z_{i;}{}^{c[b_1,b_2]}+\tfrac{1}{15}\,\varepsilon^{i[a_1a_2|[b_1}Z_{i;}{}^{b_2]c,|a_3]}-\tfrac{1}{15}\,\varepsilon^{i[b_1|[a_1a_2}Z_{i;}{}^{a_3]c,|b_2]}\nonumber\\
    &-\tfrac{1}{15}\,\varepsilon^{ic[a_1a_2}Z_{i;}{}^{a_3][b_1,b_2]}-\tfrac{1}{15}\,\varepsilon^{ib_1b_2[a_1|}Z_{i;}{}^{c|a_2,a_3]}-\tfrac{1}{15}\,\varepsilon^{ic[a_1[b_1}Z_{i;}{}^{b_2]a_2,a_3]}\label{Z2irredinverse}\;,\\
    \widehat{W}^{a_1a_2a_3a_4,c_1,c_2} :={}& -\tfrac{1}{18}\,\varepsilon^{a_1a_2a_3a_4}W^{c_1c_2}-\tfrac{1}{9}\,\varepsilon^{[a_1a_2a_3(c_1}W^{c_2)a_4]}\label{W2irredinverse}\;.
\end{align}
At the second level of higher dualisation, decomposing $D_{a_1a_2;}{}^{c_1c_2c_3,d}$
recasts the action as
\begin{equation}
    S[D_{ab;}{}^{cde,f},\widehat{Z}^{abc,d}]=S[X_{a_1a_2;}{}^{c_1c_2c_3,d},Y_{a;}{}^{c_1c_2c_3},Z_{a;}{}^{c_1c_2,d},W^{c_1c_2},\widehat{Z}^{abc,d}]\;,
\end{equation}
although \eqref{X2irred}--\eqref{W2irred} allows us to express this action as
\begin{equation}\label{dualAYZWZ}
    S[A^{a_1a_2,b_1b_2,c_1,c_2},\widehat{Y}^{a_1a_2a_3,c_1,c_2,c_3},
    \widehat{Z}^{a_1a_2a_3,b_1b_2,c},\widehat{W}^{a_1a_2a_3a_4,c_1,c_2},
    \widehat{Z}^{a_1a_2a_3,c}]\;.
\end{equation}

Off-shell dualisation from the dual graviton to the first higher dual graviton is
given by
\begin{align}
    \ytableausetup{smalltableaux}\ytableaushort{\null\null}
    \quad
    \overset{\mathcal{D}_A\:}{\longrightarrow}\quad
    & 1\times{}\ytableaushort{\null\null\null,\null}
    \quad
    \oplus\quad 1\times{}\ytableaushort{\null\null,\null,\null}
\end{align}
where $\mathcal{D}_A$ denotes one round of off-shell dualisation applied only to
the dual graviton at the previous level. At the next level, dualising only
the first higher dual graviton $A^{a_1a_2,bc}$ produces a new action 
$\mathcal{D}^2_A(S[A_{ab}])=\mathcal{D}_A(S[A^{ab,cd},\widehat{Z}^{abc,d}])$ 
with the following set of fields
\begin{align}
    \ytableausetup{smalltableaux}\ytableaushort{\null\null}
    \quad
    \overset{\mathcal{D}_A^2\:}{\longrightarrow}\quad
    & 1\times{}\ytableaushort{\null\null\null\null,\null\null}
    \oplus 1\times{}\ytableaushort{\null\null\null\null,\null,\null}
    \oplus 1\times{}\ytableaushort{\null\null\null,\null\null,\null}
    \oplus 1\times{}\ytableaushort{\null\null\null,\null,\null,\null}
    \oplus 1\times{}\ytableaushort{\null\null,\null,\null}
\end{align}
where we can see that $\widehat{Z}^{abc,d}\sim\mathbb{Y}[3,1]$ 
has been carried through to the new dual action in \eqref{dualAYZWZ} with the 
same gauge symmetries as it had in 
$S[A^{ab,cd},\widehat{Z}^{abc,d}]=\mathcal{D}_A(S[A_{ab}])\,$. 

Taking $\mathcal{D}^2_A(S[A_{ab}])$ and dualising 
only the second higher dual graviton $A^{a_1a_2,b_1b_2,c,d}$ gives us
\begin{align}
    \ytableausetup{smalltableaux}\ytableaushort{\null\null}
    \quad
    \overset{\mathcal{D}_A^3\:}{\longrightarrow}\quad
    & 1\times{}\ytableaushort{\null\null\null\null\null,\null\null\null}
    \oplus 1\times{}\ytableaushort{\null\null\null\null\null,\null\null,\null}
    \oplus 1\times{}\ytableaushort{\null\null\null\null,\null\null\null,\null}
    \oplus 1\times{}\ytableaushort{\null\null\null\null,\null\null,\null,\null}\nonumber\\
    &\oplus 1\times{}\ytableaushort{\null\null\null\null,\null,\null}
    \oplus 1\times{}\ytableaushort{\null\null\null,\null\null,\null}
    \oplus 1\times{}\ytableaushort{\null\null\null,\null,\null,\null}
    \oplus 1\times{}\ytableaushort{\null\null,\null,\null}
\end{align}
The pattern is starting to become clear now. Label groups of $k$ symmetric and $k$
antisymmetric indices by $a(k)$ and $a[k]$, respectively. For example,
$A^{a_1a_2,b_1b_2}\equiv{}A^{a[2],b(2)}$ and $\widehat{Z}^{a_1a_2a_3,b}\equiv{}\widehat{Z}^{a[3],b}$\,.
After dualising the $(n-1)^\mathrm{th}$ higher dual graviton 
$A^{a^1[2],a^2[2],\,\ldots\,,a^{n-1}[2],c(2)}\,$, 
the set of independent fields will contain the $n^{\mathrm{th}}$ higher 
dual graviton
\begin{equation}
    A^{(n)}_{[2,\ldots,2,1,1]}\equiv{}A^{(n)}
    :=A^{a^1[2],a^2[2],\,\ldots\,,a^{n-1}[2],a^n[2],c(2)}
    \quad\sim\quad
    \ytableausetup{mathmode, boxsize=1.5em}
    \begin{ytableau}
    a^1_1 & a^2_1 & \cdots & a^n_1 & c_1 & c_2 \\
    a^1_2 & a^2_2 & \cdots & a^n_2
    \end{ytableau}
\end{equation}
which is a $\mathrm{GL}(4)$-irreducible field of type 
$\mathbb{Y}[2,\ldots,2,1,1]=\mathbb{Y}(n+2,2)$\,. 
The extra fields that are produced belong to the following families
at the $n^\mathrm{th}$ level of higher dualisation:
\begin{align}
    \widehat{Y}^{(n)}_{[3,2,\ldots,2,1,1,1]}\equiv{}\widehat{Y}^{(n)}&:=\widehat{Y}^{a[3],b^1[2],\,\ldots\,,b^{n-2}[2],c(3)}
    \quad&\sim\quad
    \ytableausetup{mathmode, boxsize=1.5em}
    \begin{ytableau}
    a & b & \cdots & b & c & c & c \\
    a & b & \cdots & b \\
    a
    \end{ytableau}\\
    \widehat{Z}^{(n)}_{[3,2,\ldots,2,1]}\equiv{}\widehat{Z}^{(n)}&:=\widehat{Z}^{a[3],b^1[2],\,\ldots\,,b^{n-1}[2],c}
    \quad&\sim\quad
    \ytableausetup{mathmode, boxsize=1.5em}
    \begin{ytableau}
    a & b & \cdots & b & b & c \\
    a & b & \cdots & b & b \\
    a
    \end{ytableau}\quad\;\;\\
    \widehat{W}^{(n)}_{[4,2,\ldots,2,1,1]}\equiv{}\widehat{W}^{(n)}&:=\widehat{W}^{a[4],b^1[2],\,\ldots\,,b^{n-2}[2],c(2)}
    \quad&\sim\quad
    \ytableausetup{mathmode, boxsize=1.5em}
    \begin{ytableau}
    a & b & \cdots & b & c & c & \none \\
    a & b & \cdots & b \\
    a \\
    a
    \end{ytableau}
\end{align}
As with \eqref{antisymlevel2X}--\eqref{antisymlevel2W} in Appendix
\ref{appendix-phipsi}, these irreducible fields arise, respectively, from fields
\begin{equation}
    \phi^{a(n+2),b(n)}\;,\qquad
    \psi_{_Y}^{a(n+2),b(n-1),c}\;,\qquad
    \psi_{_Z}^{a(n+1),b(n),c}\;,\qquad
    \psi_{_W}^{a(n+1),b(n-1),c,d}\;,
\end{equation}
in the symmetric convention. They themselves arise from the traceless components
of the dual field $D_{a_1a_2;}{}^{c(n+2),d(n)}$ that must be introduced when moving
from the $(n-1)^\mathrm{th}$ to the $n^\mathrm{th}$ level of higher dualisation. 
This is a result of the Young diagram decomposition
\begin{align}
    \ytableausetup{smalltableaux}
    \ytableaushort{\null,\null}
    \quad\otimes\quad \ytableaushort{\null\cdots\null\null\null,\null\cdots\null}
    \quad
    \sim
    \quad
    &
    \ytableaushort{\null\cdots\null\null\null\null,\null\cdots\null\null}
    \hspace{0.2cm}\oplus\hspace{0.2cm}
    \ytableaushort{\null\cdots\null\null\null\null,\null\cdots\null,\null}
    \hspace{0.2cm}\oplus\hspace{0.2cm}
    \ytableaushort{\null\cdots\null\null\null,\null\cdots\null\null,\null}
    \hspace{0.2cm} \oplus\hspace{0.2cm} 
    \ytableaushort{\null\cdots\null\null\null,\null\cdots\null,\null,\null}
    \label{nthlevelforA}
\end{align}
which generalises \eqref{secondlevelforA}. 
Ultimately, at the $n^\mathrm{th}$ level of higher dualisation, 
the action will be given in terms of the following set of independent fields:
\begin{equation}
    \left\{A^{(n)},\widehat{Y}^{(n)},\widehat{Z}^{(n)},\widehat{W}^{(n)}\right\}
    \cup
    \left\{\widehat{Y}^{(n-1)},\widehat{Z}^{(n-1)},\widehat{W}^{(n-1)}\right\}
    \cup
    \cdots
    \cup
    \left\{\widehat{Y}^{(2)},\widehat{Z}^{(2)},\widehat{W}^{(2)}\right\}
    \cup
    \left\{\widehat{Z}^{(1)}\right\}
\end{equation}

In parallel with the $A^{(n)}$ family of dual graviton fields, the $\widehat{Z}^{(n)}$ family
of extra fields starts to appear at the first level of higher dualisation when 
$\widehat{Z}^{abc,d}$ enters the action, while the $\widehat{Y}^{(n)}$ and
$\widehat{W}^{(n)}$ families both enter the action at the second level. With index structure explicit, we see that there is only $A_{ab}\sim A^{(0)}_{[1,1]}$
at level zero, i.e. at the level of the usual dual graviton.

\paragraph{Dualisation at low levels.}

Up to this point, we have only dualised the $A^{(n)}$ family of fields and the extra
fields have been completely untouched so that they are carried forward into every
new dual action. This dualisation scheme may be extended by dualising
some or all of the extra fields that we encounter at each stage.
In this case, the second level of higher dualisation for the first higher dual
graviton $A^{ab,cd}$ and the extra field $\widehat{Z}^{abc,d}$ is summarised as
\begin{align}
    \left\{A^{(1)}_{[2,1,1]}\right\}&\overset{\mathcal{D}}{\longrightarrow}
    \left\{A^{(2)}_{[2,2,1,1]},
    \widehat{Y}^{(2)}_{[3,1,1,1]},
    \widehat{Z}^{(2)}_{[3,2,1]},
    \widehat{W}^{(2)}_{[4,1,1]}\right\}\;,
    \label{dualiseX1}\\
    \left\{\widehat{Z}^{(1)}_{[3,1]}\right\}&\overset{\mathcal{D}}{\longrightarrow}
    \left\{\widehat{Z}^{(2)}_{[3,2,1]},
    \widehat{W}^{(2)}_{[4,1,1]},
    \widehat{P}^{(2)}_{[4,2]},
    \widehat{Q}^{(2)}_{[3,3]}\right\}\;
    \label{dualiseZ1},
\end{align}
or equivalently as
\begin{align}
    \left\{A^{(0)}\right\}&\overset{\mathcal{D}^2}{\longrightarrow}
    \left\{1\times{}A^{(2)},
    1\times{}\widehat{Y}^{(2)},
    2\times{}\widehat{Z}^{(2)},
    2\times{}\widehat{W}^{(2)},
    1\times{}\widehat{P}^{(2)},
    1\times{}\widehat{Q}^{(2)}\right\}\label{dualiselevel2}\;,
\end{align}
where $\mathcal{D}$ denotes one round of off-shell dualisation applied to
\emph{every} field at the previous level, not just the $A^{(n)}$ field. Moreover,
$\widehat{P}^{(n)}$ and $\widehat{Q}^{(n)}$ denote two new families of fields with
indices grouped as $\mathbb{Y}[4,2,\ldots,2]$ and $\mathbb{Y}[3,3,2,\ldots,2]$\,, 
respectively, which only appear after $\widehat{Z}^{abc,d}$ has been dualised.
To further illustrate the ever growing number of families of fields, the third level
of higher dualisation can be summarised as
\begin{align}
    \left\{A^{(2)}_{[2,2,1,1]}\right\}&\overset{\mathcal{D}}{\longrightarrow}
    \left\{A^{(3)}_{[2,2,2,1,1]},
    \widehat{Y}^{(3)}_{[3,2,1,1,1]},
    \widehat{Z}^{(3)}_{[3,2,2,1]},
    \widehat{W}^{(3)}_{[4,2,1,1]}\right\}\;,\\
    \left\{\widehat{Y}^{(2)}_{[3,1,1,1]}\right\}&\overset{\mathcal{D}}{\longrightarrow}
    \left\{\widehat{Y}^{(3)}_{[3,2,1,1,1]},
    \widehat{W}^{(3)}_{[4,2,1,1]},
    \widehat{R}^{(3)}_{[4,1,1,1,1]},
    \widehat{S}^{(3)}_{[3,3,1,1]}\right\}\;,\\
    \left\{\widehat{Z}^{(2)}_{[3,2,1]}\right\}&\overset{\mathcal{D}}{\longrightarrow}
    \left\{\widehat{Z}^{(3)}_{[3,2,2,1]},
    \widehat{W}^{(3)}_{[4,2,1,1]},
    \widehat{P}^{(3)}_{[4,2,2]},
    \widehat{Q}^{(3)}_{[3,3,2]},
    \widehat{S}^{(3)}_{[3,3,1,1]},
    \widehat{T}^{(3)}_{[4,3,1]}\right\}\;,\\
    \left\{\widehat{W}^{(2)}_{[4,1,1]}\right\}&\overset{\mathcal{D}}{\longrightarrow}
    \left\{\widehat{W}^{(3)}_{[4,2,1,1]}, \widehat{T}^{(3)}_{[4,3,1]}\right\}\;,\\
    \left\{\widehat{P}^{(2)}_{[4,2]}\right\}&\overset{\mathcal{D}}{\longrightarrow}
    \left\{\widehat{P}^{(3)}_{[4,2,2]}, \widehat{T}^{(3)}_{[4,3,1]}, \widehat{O}^{(3)}_{[4,4]}\right\}\;,\\
    \left\{\widehat{Q}^{(2)}_{[3,3]}\right\}&\overset{\mathcal{D}}{\longrightarrow}
    \left\{\widehat{Q}^{(3)}_{[3,3,2]}, \widehat{T}^{(3)}_{[4,3,1]}\right\}\;,
\end{align}
or equivalently as
\begin{align}
    \Big\{A^{(0)}\Big\}
    \overset{\mathcal{D}^3}{\longrightarrow}\Big\{&1\times{}A^{(3)},2\times{}\widehat{Y}^{(3)},3\times{}\widehat{Z}^{(3)},6\times{}\widehat{W}^{(3)},3\times{}\widehat{P}^{(3)},
    \nonumber\\
    &
    3\times{}\widehat{Q}^{(3)},1\times{}\widehat{R}^{(3)},3\times{}\widehat{S}^{(3)},6\times{}\widehat{T}^{(3)},1\times{}\widehat{O}^{(3)}\Big\}\;.\label{dualiselevel3}
\end{align}
Let's take inventory. 
At levels one and two, dualising every field at every level, we have
\begin{align}
    \ytableausetup{smalltableaux}\ytableaushort{\null\null}
    \quad
    \overset{\mathcal{D}}{\longrightarrow}\quad
    & 1\times{}\ytableaushort{\null\null\null,\null}
    \quad
    \oplus\quad 1\times{}\ytableaushort{\null\null,\null,\null}\\
    \ytableausetup{smalltableaux}\ytableaushort{\null\null}
    \quad
    \overset{\mathcal{D}^2}{\longrightarrow}\quad
    & 1\times{}\ytableaushort{\null\null\null\null,\null\null}
    \oplus 1\times{}\ytableaushort{\null\null\null\null,\null,\null}
    \oplus 2\times{}\ytableaushort{\null\null\null,\null\null,\null}
    \oplus 2\times{}\ytableaushort{\null\null\null,\null,\null,\null}
    \oplus 1\times{}\ytableaushort{\null\null,\null\null,\null,\null}
    \oplus 1\times{}\ytableaushort{\null\null,\null\null,\null\null}\label{dualiselevel2YT}
\end{align}
%
%
whereas the third level is visualised as
\begin{align}
    \ytableausetup{smalltableaux}\ytableaushort{\null\null}
    \quad
    \overset{\mathcal{D}^3}{\longrightarrow}\quad
    &
    1\times{}\ytableaushort{\null\null\null\null\null,\null\null\null}
    \oplus
    2\times{}\ytableaushort{\null\null\null\null\null,\null\null,\null}
    \oplus
    3\times{}\ytableaushort{\null\null\null\null,\null\null\null,\null}
    \oplus
    6\times{}\ytableaushort{\null\null\null\null,\null\null,\null,\null}
    \oplus
    3\times{}\ytableaushort{\null\null\null,\null\null\null,\null,\null}
    \nonumber\\
    &\oplus
    3\times{}\ytableaushort{\null\null\null,\null\null\null,\null\null}
    \oplus
    1\times{}\ytableaushort{\null\null\null\null\null,\null,\null,\null}
    \oplus
    3\times{}\ytableaushort{\null\null\null\null,\null\null,\null\null}
    \oplus
    6\times{}\ytableaushort{\null\null\null,\null\null,\null\null,\null}
    \oplus
    1\times{}\ytableaushort{\null\null,\null\null,\null\null,\null\null}\label{dualiselevel3YT}
\end{align}
The third higher dual graviton $A^{(3)}\equiv{}A^{a_1a_2,b_1b_2,c_1c_2,(de)}$ dualises off-shell to produce
\begin{align}
    \mathbb{Y}[2,2,2,1,1]
    \longrightarrow
    \null&\mathbb{Y}[2,2,2,2,1,1]\oplus\mathbb{Y}[3,2,2,2,1]\oplus\mathbb{Y}[3,2,2,1,1,1]\oplus\mathbb{Y}[4,2,2,1,1]\nonumber
\end{align}
which consists of the fields $A^{(4)}$, $\widehat{Y}^{(4)}$, $\widehat{Z}^{(4)}$ and $\widehat{W}^{(4)}$. In addition to this, the extra fields at the third level of higher dualisation can also be dualised off-shell, and they produce the following fields at the fourth level of higher dualisation:
\begin{align}
    \mathbb{Y}[3,2,1,1,1]
    \longrightarrow
    \null&\mathbb{Y}[4,3,1,1,1]\oplus\mathbb{Y}[4,2,2,1,1]\oplus\mathbb{Y}[4,2,1,1,1,1]\oplus\mathbb{Y}[3,3,2,1,1]\nonumber\\
    &\oplus\mathbb{Y}[3,3,1,1,1,1]\oplus\mathbb{Y}[3,2,2,1,1,1]\nonumber\\
    \mathbb{Y}[3,2,2,1]
    \longrightarrow
    \null&\mathbb{Y}[4,3,2,1]\oplus\mathbb{Y}[4,2,2,2]\oplus\mathbb{Y}[4,2,2,1,1]\oplus\mathbb{Y}[3,3,2,2]\nonumber\\
    &\oplus\mathbb{Y}[3,3,2,1,1]\oplus\mathbb{Y}[3,2,2,2,1]\nonumber\\
    \mathbb{Y}[4,2,1,1]
    \longrightarrow
    \null&\mathbb{Y}[4,2,2,1,1]\oplus\mathbb{Y}[4,3,2,1]\oplus\mathbb{Y}[4,3,1,1,1]\oplus\mathbb{Y}[4,4,1,1]\nonumber\\
    \mathbb{Y}[4,2,2]
    \longrightarrow
    \null&\mathbb{Y}[4,4,2]\oplus\mathbb{Y}[4,2,2,2]\oplus\mathbb{Y}[4,3,2,1]\nonumber\\
    \mathbb{Y}[3,3,2]
    \longrightarrow
    \null&\mathbb{Y}[4,3,3]\oplus\mathbb{Y}[4,3,2,1]\oplus\mathbb{Y}[3,3,3,1]\oplus\mathbb{Y}[3,3,2,2]\nonumber\\
    \mathbb{Y}[4,1,1,1,1]
    \longrightarrow
    \null&\mathbb{Y}[4,3,1,1,1]\oplus\mathbb{Y}[4,2,1,1,1,1]\nonumber\\
    \mathbb{Y}[3,3,1,1]
    \longrightarrow
    \null&\mathbb{Y}[4,3,2,1]\oplus\mathbb{Y}[4,3,1,1,1]\oplus\mathbb{Y}[3,3,3,1]\oplus\mathbb{Y}[3,3,2,1,1]\nonumber\\
    \mathbb{Y}[4,3,1]
    \longrightarrow
    \null&\mathbb{Y}[4,4,2]\oplus\mathbb{Y}[4,4,1,1]\oplus\mathbb{Y}[4,3,3]\oplus\mathbb{Y}[4,3,2,1]\nonumber\\
    \mathbb{Y}[4,4]
    \longrightarrow
    \null&\mathbb{Y}[4,4,2]\nonumber
\end{align}
The irreducible fields at level four are given in column notation as
\begin{gather}
    \mathbb{Y}[2,2,2,2,1,1]\;,\quad\mathbb{Y}[3,2,2,2,1]\;,\quad\mathbb{Y}[3,2,2,1,1,1]\;,\quad\mathbb{Y}[4,2,2,1,1]\;,\quad\mathbb{Y}[4,3,1,1,1],\nonumber\\
    \mathbb{Y}[4,2,1,1,1,1]\;,\quad\mathbb{Y}[3,3,2,1,1]\;,\quad\mathbb{Y}[3,3,1,1,1,1]\;,\quad\mathbb{Y}[4,3,2,1]\;,\quad\mathbb{Y}[4,2,2,2]\;,\quad\nonumber\\
    \mathbb{Y}[3,3,2,2]\;,\quad\mathbb{Y}[4,4,1,1]\;,\quad\mathbb{Y}[4,4,2]\;,\quad\mathbb{Y}[4,3,3]\;,\quad\mathbb{Y}[3,3,3,1]
\end{gather}
with corresponding Young diagrams
\begin{gather}
    \ytableaushort{\null\null\null\null\null\null,\null\null\null\null}
    \quad \ytableaushort{\null\null\null\null\null,\null\null\null\null,\null}
    \quad \ytableaushort{\null\null\null\null\null\null,\null\null\null,\null}
    \quad \ytableaushort{\null\null\null\null\null,\null\null\null,\null,\null}
    \quad
    \ytableaushort{\null\null\null\null\null,\null\null,\null\null,\null}
    \quad
    \ytableaushort{\null\null\null\null\null\null,\null\null,\null,\null}
    \quad \ytableaushort{\null\null\null\null\null,\null\null\null,\null\null}\nonumber\\
    \ytableaushort{\null\null\null\null\null\null,\null\null,\null\null}
    \quad
    \ytableaushort{\null\null\null\null,\null\null\null,\null\null,\null}
    \quad \ytableaushort{\null\null\null\null,\null\null\null\null,\null,\null}
    \quad
    \ytableaushort{\null\null\null\null,\null\null\null\null,\null\null}
    \quad
    \ytableaushort{\null\null\null\null,\null\null,\null\null,\null\null}
    \quad
    \ytableaushort{\null\null\null,\null\null\null,\null\null,\null\null}
    \quad
    \ytableaushort{\null\null\null,\null\null\null,\null\null\null,\null}
    \quad
    \ytableaushort{\null\null\null\null,\null\null\null,\null\null\null}
\end{gather}
and respective multiplicities in the order presented above
\begin{align}
    (1\,,\;4\,,\;3\,,\;12\,,\;12\,,\;3\,,\;8\,,\;2\,,\;24\,,\;
    6\,,\;6\,,\;12\,,\;10\,,\;9\,,\;6)\;.
\end{align}
The number of families of fields will clearly continue to increase with further dualisation.

Finally, note that $\widehat{W}^{(n)}$ is the same as $A^{(n-2)}$ with four extra 
antisymmetric indices, so they are Hodge dual. Dualising every field at every stage
but only keeping track of the $A^{(n)}$ and $\widehat{W}^{(n)}$ families, we know
that off-shell dualisation of $A^{(n)}$ produces both $A^{(n+1)}$ and
$\widehat{W}^{(n+1)}=*A^{(n-1)}\,$. At the $n^\mathrm{th}$ level of higher dualisation,
we have one copy of $A^{(n)}$ and at least one copy of the $k^\mathrm{th}$ Hodge dual
$A^{(n-2k)}$ for every positive integer $k$ such that $0\leq{}n-2k\leq{}n\,$. 
When $n$ is even, we find that our set of independent fields contains
$\left\{A^{(0)},A^{(2)},A^{(4)},\ldots,A^{(n)}\right\}$ as a subset. 
Similarly, when $n$ is odd, we find that it contains
$\left\{A^{(1)},A^{(3)},\ldots,A^{(n)}\right\}$ instead. However, if we only
dualise the $A^{(n)}$ fields, and if we make the appropriate correspondence
between the $W^{(n)}$ and $A^{(n)}$ families, then we find
$\left\{A^{(n)}\right\}\cup\left\{A^{(n-2)},A^{(n-3)},\ldots,A^{(1)},A^{(0)}\right\}$ at level $n$\,.

\paragraph{Summary.} 
Off-shell dualisation on empty columns in the Young tableaux of every field at the $n^\mathrm{th}$ level of higher dualisation produces independent fields in a 
one-to-one correspondance with the set of $\mathrm{GL}(4)$-irreducible fields
entering the decomposition of the tensor product
\begin{align}
    \ytableausetup{smalltableaux}
    \ytableaushort{\null,\null}\quad\otimes\quad
    \ytableaushort{\null,\null}\quad\otimes\quad \ldots
    \quad\otimes\quad 
    \ytableaushort{\null,\null}\quad\otimes\quad  \ytableaushort{\null\null}
\end{align}
with $n$ factors of the antisymmetric Young diagram 
$~\ytableausetup{smalltableaux}\ytableaushort{\null,\null}~\,$.

\subsection{Contact with $A_1^{+++}$}
\label{section:contact}

Before we explain the correspondence between the $\ell_2$ representation of 
$A_1^{+++}$ and the extra fields produced via off-shell dualisation, it will be
useful to review an efficient method for computing fundamental representations 
\cite{Kleinschmidt:2003jf}. First of all, add a new node labelled $*$ to the
Dynkin diagram of $A_1^{+++}$ and attach it to node $i$, say, by a single edge.
A generic root in the corresponding enlarged algebra $A_1^{+++(i)}$ is given by
\begin{equation}
    \alpha=m_*\alpha_*+m_4\alpha_4+\sum_{j=1}^{3}m_j\alpha_j
\end{equation}
where $\alpha_*$ denotes the new simple root associated to the new node $*$ and
where $m_4$ is the level in the usual decomposition of $A_1^{+++}$ with respect
to its $A_3$ subalgebra. The structure of $A_1^{+++}$ is studied at each level
by looking at the representation content (i.e. the weight space) of the $A_3$
subalgebra. Any generic $A_3$ weight can be expressed as
\begin{equation}
    \lambda=\sum_{i=1}^{3}p_i\lambda_i
\end{equation}
where $\lambda_i$ is the $i^\mathrm{th}$ fundamental weight of $A_3$\,.
This weight may also be written as $\lambda=[p_1,p_2,p_3]$ and we may depict this
weight (and its corresponding representation) by a Young diagram with $p_3$
columns of height 1, $p_2$ columns of height 2, and $p_1$ columns of height 3.
With this, we have
\begin{equation}
    \lambda=[p_1,p_2,p_3]
    \sim\mathbb{Y}[\underbrace{3,\ldots,3}_{p_1},
    \underbrace{2,\ldots,2}_{p_2},
    \underbrace{1,\ldots,1}_{p_3}]
    =:\mathbb{Y}[3^{p_1},2^{p_2},1^{p_3}]\;.
\end{equation}
The relationship between the permitted $A_3$ Dynkin labels $p_i$ of $\lambda$
and the Kac labels $m_i$ of the $A_1^{+++}$ root associated with $\lambda$ can
be found in equation (16.6.3) of reference \cite{West:2012vka}.

The notion of level is preserved by commutators, so the set of roots with $m_*=1$
forms a representation of $A_1^{+++}$ which one can show is equivalent to the
$i^\mathrm{\,th}$ fundamental representation, denoted $\ell_i$\,.
The $\ell_1$ and $\ell_2$ representations of $A_1^{+++}$ were calculated this way
in the tables found in this paper. Note that the $A_3$ weights in the tables for
$\ell_1$ and $\ell_2$ have their corresponding $A_1^{+++}$ roots written as
$A_1^{+++(i)}$ roots so that the new simple root $\alpha_*$ is included.

In previous sections, we have found the extra fields appearing in the action
principles and duality relations at low levels. Now we are finally ready to
show, level-by-level, that off-shell dualisation produces a set of extra fields
that is closely correlated with the $\ell_2$ representation. In particular, at the
$n^\mathrm{th}$ level of higher dualisation, we count fields that appear
in the adjoint representation at level $n+1$ and in the $\ell_2$ representation
at level $n$. This will then be compared against the set of extra fields 
that we obtain by off-shell dualising every field at every level.

In Table \ref{table:comparison}, the `adj' and `$\ell_2$' columns contain
the field multiplicities in the adjoint and $\ell_2$ representations,
respectively, and the `total' column gives their sum. The `maximal off-shell'
column tells us how many of each field is found by off-shell dualising every
field at every level. Lastly, the `net' column is the `maximal off-shell'
column minus the `total' column. It tells us if we have too many, too few,
or the right amount of fields with maximal off-shell dualisation.

\begin{table}[h]
\caption{The adjoint representation of $A_1^{+++}$ up to level four.}
\centering
\begin{tabular}{|c|c|c|c|c|c|c|}\hline
$l$&$A_{3}$ weight&$A_1^{+++}$ root $\alpha$&$\alpha^2$&mult.&field&label\\\hline\hline
$0$&$[1,0,1]$&$(1,1,1,0)$&$2$&$1$&$h_a{}^b$&$h$\\\hline
$1$&$[0,0,2]$&$(0,0,0,1)$&$2$&$1$&$A^{(0)}_{[1,1]}$&$a_0$\\\hline
$2$&$[0,1,2]$&$(0,0,1,2)$&$2$&$1$&$A^{(1)}_{[2,1,1]}$&$a_1$\\\hline
$3$&$[0,2,2]$&$(0,0,2,3)$&$2$&$1$&$A^{(2)}_{[2,2,1,1]}$&$a_2$\\
$3$&$[1,1,1]$&$(0,1,3,3)$&$-4$&$1$&$\widehat{Z}^{(2)}_{[3,2,1]}$&$c_2$\\\hline
$4$&$[0,3,2]$&$(0,0,3,4)$&$2$&$1$&$A^{(3)}_{[2,2,2,1,1]}$&$a_3$\\
$4$&$[1,1,3]$&$(0,1,3,4)$&$-2$&$1$&$\widehat{Y}^{(3)}_{[3,2,1,1,1]}$&$d_1$\\
$4$&$[1,2,1]$&$(0,1,4,4)$&$-6$&$2$&$\widehat{Z}^{(3)}_{[3,2,2,1]}$&$d_2$\\
$4$&$[0,1,2]$&$(1,2,4,4)$&$-10$&$1$&$\widehat{W}^{(3)}_{[4,2,1,1]}$&$d_3$\\
$4$&$[2,1,0]$&$(0,2,5,4)$&$-10$&$1$&$\widehat{Q}^{(3)}_{[3,3,2]}$&$d_5$\\
$4$&$[2,0,2]$&$(0,2,4,4)$&$-8$&$1$&$\widehat{S}^{(3)}_{[3,3,1,1]}$&$d_7$\\
$4$&$[1,0,1]$&$(1,3,5,4)$&$-14$&$1$&$\widehat{T}^{(3)}_{[4,3,1]}$&$d_8$\\\hline
\end{tabular}
\label{A1+++-adj-level01234}
\end{table}

\begin{table}[h]
\caption{The $\ell_2$ representation of $A_1^{+++}$ up to level three.}
\centering
\begin{tabular}{|c|c|c|c|c|c|c|}\hline
$l$&$A_{3}$ weight&$A_1^{+++(2)}$ root $\alpha$&$\alpha^2$&mult.&field&label\\\hline\hline
$0$&$[0,1,0]$&$(0,0,0,0,1)$&$2$&$1$&$\widehat{U}^{(0)}_{[2]}$&$u$\\\hline
$1$&$[1,0,1]$&$(0,1,1,1,1)$&$0$&$1$&$\widehat{Z}^{(1)}_{[3,1]}$&$b$\\\hline
$2$&$[1,0,3]$&$(0,1,1,2,1)$&$2$&$1$&$\widehat{Y}^{(2)}_{[3,1,1,1]}$&$c_1$\\
$2$&$[1,1,1]$&$(0,1,2,2,1)$&$-2$&$1$&$\widehat{Z}^{(2)}_{[3,2,1]}$&$c_2$\\
$2$&$[0,0,2]$&$(1,2,2,2,1)$&$-4$&$1$&$\widehat{W}^{(2)}_{[4,1,1]}$&$c_3$\\
$2$&$[0,1,0]$&$(1,2,3,2,1)$&$-6$&$1$&$\widehat{P}^{(2)}_{[4,2]}$&$c_4$\\\hline
$3$&$[1,1,3]$&$(0,1,2,3,1)$&$0$&$1$&$\widehat{Y}^{(3)}_{[3,2,1,1,1]}$&$d_1$\\
$3$&$[1,2,1]$&$(0,1,3,3,1)$&$-4$&$2$&$\widehat{Z}^{(3)}_{[3,2,2,1]}$&$d_2$\\
$3$&$[0,1,2]$&$(1,2,3,3,1)$&$-8$&$4$&$\widehat{W}^{(3)}_{[4,2,1,1]}$&$d_3$\\
$3$&$[0,2,0]$&$(1,2,4,3,1)$&$-10$&$2$&$\widehat{P}^{(3)}_{[4,2,2]}$&$d_4$\\
$3$&$[2,1,0]$&$(0,2,4,3,1)$&$-8$&$1$&$\widehat{Q}^{(3)}_{[3,3,2]}$&$d_5$\\
$3$&$[0,0,4]$&$(1,2,2,3,1)$&$-2$&$1$&$\widehat{R}^{(3)}_{[4,1,1,1,1]}$&$d_6$\\
$3$&$[2,0,2]$&$(0,2,3,3,1)$&$-6$&$1$&$\widehat{S}^{(3)}_{[3,3,1,1]}$&$d_7$\\
$3$&$[1,0,1]$&$(1,3,4,3,1)$&$-12$&$2$&$\widehat{T}^{(3)}_{[4,3,1]}$&$d_8$\\\hline
\end{tabular}
\label{A1+++-l2-level0123}
\end{table}

\begin{table}[h]
\caption{Extra fields from off-shell dualisation compared with $A_1^{+++}$ representations.}
\centering
\begin{tabular}{|c|c|c|c|c|c|c|}\hline
label&$A_3$ weight&adj&$\ell_2$&total&maximal off-shell&net\\\hline\hline
$b$&$[1,0,1]$&$0$&$1$&$1$&$1$&$0$\\\hline
$c_1$&$[1,0,3]$&$0$&$1$&$1$&$1$&$0$\\
$c_2$&$[1,1,1]$&$1$&$1$&$2$&$2$&$0$\\
$c_3$&$[0,0,2]$&$0$&$1$&$1$&$2$&$+1$\\
$c_4$&$[0,1,0]$&$0$&$1$&$1$&$1$&$0$\\
$c_5$&$[2,0,0]$&$0$&$0$&$0$&$1$&$+1$\\\hline
$d_1$&$[1,1,3]$&$1$&$1$&$2$&$2$&$0$\\
$d_2$&$[1,2,1]$&$2$&$2$&$4$&$3$&$-1$\\
$d_3$&$[0,1,2]$&$1$&$4$&$5$&$6$&$+1$\\
$d_4$&$[0,2,0]$&$0$&$2$&$2$&$3$&$+1$\\
$d_5$&$[2,1,0]$&$1$&$1$&$2$&$3$&$+1$\\
$d_6$&$[0,0,4]$&$0$&$1$&$1$&$1$&$0$\\
$d_7$&$[2,0,2]$&$1$&$1$&$2$&$3$&$+1$\\
$d_8$&$[1,0,1]$&$1$&$2$&$3$&$6$&$+3$\\
$d_9$&$[0,0,0]$&$0$&$0$&$0$&$1$&$+1$\\\hline
\end{tabular}
\label{table:comparison}
\end{table}

\begin{table}[h]
\caption{The $\ell_1$ representation of $A_1^{+++}$ up to level three.}
\centering
\begin{tabular}{|c|c|c|c|c|c|}\hline
$l$&$A_{3}$ weight&$A_1^{+++(1)}$ root $\alpha$&$\alpha^2$&mult.&field\\\hline\hline
$0$&$[1,0,0]$&$(0,0,0,0,1)$&$2$&$1$&$P^{\,a}$\\\hline
$1$&$[0,0,1]$&$(1,1,1,1,1)$&$0$&$1$&$Z_{[1]}$\\\hline
$2$&$[0,1,1]$&$(1,1,2,2,1)$&$-2$&$1$&$Z_{[2,1]}$\\
$2$&$[0,0,3]$&$(1,1,1,2,1)$&$2$&$1$&$Z_{[1,1,1]}$\\\hline
$3$&$[1,1,0]$&$(1,2,4,3,1)$&$-8$&$1$&$Z_{[3,2]}$\\
$3$&$[1,0,2]$&$(1,2,3,3,1)$&$-6$&$1$&$Z_{[3,1,1]}$\\
$3$&$[0,2,1]$&$(1,1,3,3,1)$&$-4$&$2$&$Z_{[2,2,1]}$\\
$3$&$[0,1,3]$&$(1,1,2,3,1)$&$0$&$1$&$Z_{[2,1,1,1]}$\\\hline
\end{tabular}
\label{A1+++-l1-level0123}
\end{table}

\paragraph{From gravity to dual gravity.}

Recall the dualisation of gravity in Section \ref{section:grav2dualgrav} which
gave us the dual graviton with $\mathrm{GL}(4)$-irreducible symmetry type $\mathbb{Y}[D-3,1]$
and an extra $(D-2)$-form which may be gauged away with a shift symmetry
\cite{West:2001as}. In four dimensions, the decomposition
\begin{equation}
    \ytableaushort{\null}~\otimes~\ytableaushort{\null}
    \quad=\quad
    \ytableaushort{\null\null}~\oplus~\ytableaushort{\null,\null}
\end{equation}
gives us a symmetric rank-2 field $A_{ab}$ and a 2-form field.
The symmetric field is the familiar dual graviton $a_0:=A^{(0)}_{[1,1]}$ in the
adjoint representation of $A_1^{+++}$ at level 1 in Table \ref{A1+++-adj-level01234}, 
and the 2-form $u:=\widehat{U}^{(0)}_{[2]}$ is found in the $\ell_2$ representation
at level 0 in Table \ref{A1+++-l2-level0123}. In contrast to what happens later, we do not dualise the extra 2-form field as it can be shifted away.

\paragraph{The first level of higher dualisation.}

Taking the dual graviton $A^{(0)}_{[1,1]}$ and dualising again, we obtain the
first higher dual graviton $A^{(1)}_{[2,1,1]}$ and the extra field 
$\widehat{Z}^{(1)}_{[3,1]}$\,, with symmetry types $\mathbb{Y}[2,1,1]$
and $\mathbb{Y}[3,1]$. We will now find each of these fields in the
representations of $A_1^{+++}$\,.

The weight $[0,1,2]$ in the adjoint representation at level 2 
(see Table \ref{A1+++-adj-level01234}) corresponds to the
first higher dual graviton field $a_1:=A^{(1)}_{[2,1,1]}$\,, and the weight $[1,0,1]$
in the $\ell_2$ representation at level 1 (see Table \ref{A1+++-l2-level0123})
corresponds to the extra field 
$b:=\widehat{Z}^{(1)}_{[3,1]}$ required to build a consistent dual action principle.
So, at the first level of higher dualisation, we see the complete correspondence between
the fields produced during off-shell dualisation and the generators of the adjoint and
$\ell_2$ representations of $A_1^{+++}$. There is a perfect match at this level.
In Table \ref{table:comparison}, we see that $b$ appears zero times in the adjoint
at level 2 and once at level 1 in $\ell_2$\,. It is also required exactly once at this
off-shell dualisation at this level, hence the zero in the `net' column.
Note that, although they both have the same $A_3$ weight
$[1,0,1]$, the two fields $b$ and $d_8$ in $\ell_2$ at levels 1 and 3, respectively,
are not compared or counted together since they appear at different levels.
In fact, $d_8$ should be viewed as a $\mathbb{Y}[4,3,1]$ field.

\paragraph{The second level of higher dualisation.}

Taking the first higher dual graviton $A^{(1)}_{[2,1,1]}$ and dualising again,
we obtain the second higher dual graviton $A^{(2)}_{[2,2,1,1]}$ and three
extra fields that are required for a consistent action principle:
$\widehat{Y}^{(2)}_{[3,1,1,1]}$, $\widehat{Z}^{(2)}_{[3,2,1]}$ and
$\widehat{W}^{(2)}_{[4,1,1]}$. This is the \emph{minimal off-shell dualisation}
of linearised gravity. In addition, we may dualise the extra field
from the previous level for \emph{maximal off-shell dualisation}.
Recall \eqref{dualiseX1} and \eqref{dualiseZ1} for convenience:
\begin{align}
    A^{(1)}_{[2,1,1]}
    \quad\longmapsto\quad{}&
    A^{(2)}_{[2,2,1,1]}
    ~\oplus~\widehat{Y}^{(2)}_{[3,1,1,1]}
    ~\oplus~\widehat{Z}^{(2)}_{[3,2,1]}
    ~\oplus~\widehat{W}^{(2)}_{[4,1,1]}\\
    \widehat{Z}^{(1)}_{[3,1]}
    \quad\longmapsto\quad{}&
    \widehat{Z}^{(2)}_{[3,2,1]}
    ~\oplus~\widehat{W}^{(2)}_{[4,1,1]}
    ~\oplus~\widehat{P}^{(2)}_{[4,2]}
    ~\oplus~\widehat{Q}^{(2)}_{[3,3]}    
\end{align}
We are now going to see where these fields are in the
representations of $A_1^{+++}$\,.

The weight $[0,2,2]$ in the adjoint representation at level 3 corresponds to the
second higher dual graviton $a_2:=A^{(1)}_{[2,1,1]}$\,. It is useful to look at a
couple of extra fields in detail. In the $\ell_2$ representation at level 2, we find the $A_3$
weight $[1,0,3]$ which corresponds to $c_1:=\widehat{Y}^{(2)}_{[3,1,1,1]}$\,.
Since it does not appear in the adjoint, $c_1$ appears only once in our tables.
Moreover, $c_1$ appears precisely once in maximal off-shell
dualisation at this level, so there is a perfect match for $c_1$\,.
Moving onto the next field, we see in Table \ref{table:comparison} that 
$c_2:=\widehat{Z}^{(2)}_{[3,2,1]}$ appears once in the adjoint and once in the
$\ell_2$\,, and it also appears twice in maximal off-shell dualisation: once from
the dualisation of $a_1$ and once more from the dualisation of $b$\,. Another
perfect match. The reader might like to check that $c_4:=\widehat{P}^{(2)}_{[4,2]}$
gives yet another match.

Unfortunately, we do not find a match for every extra field. For example,
$c_3:=\widehat{W}^{(2)}_{[4,1,1]}$ does not appear in the adjoint, and it appears
once in the $\ell_2$\,. However, two of them are required for maximal off-shell
dualisation. In other words, although $\ell_2$ contains enough for the minimal
off-shell description, we go slightly over when we dualise every field at every
level. It is even more peculiar with $c_5:=\widehat{Q}^{(2)}_{[3,3]}$ since it
is not contained in the tables at all, yet it is needed for maximal off-shell
dualisation. We suggest that $c_5$ should be thought of as thrice Hodge
dual to the extra field $c_3$\,.

There is a perfect match for the $\mathrm{GL}(D)$ \emph{types} of fields
required but, as for multiplicities, we appear to be lacking a small number of
fields in the tables. One possible solution could be to dualise some extra fields
but not all of them. By carefully selecting which fields to dualise, this would
provide an off-shell description of gravity that does not exceed the field content
of the adjoint and $\ell_2$ representations of $A_1^{+++}$ but nonetheless, within
this restriction, as many fields as possible are dualised.
This lies somewhere between the minimal and maximal off-shell 
dualisations, and we call it the \emph{optimal off-shell dualisation}
of linearised gravity.

\paragraph{The third level of higher dualisation.}

Moving onto the next level, we find that maximal off-shell dualisation
produces the fields in \eqref{dualiselevel3} with Young tableaux 
\eqref{dualiselevel3YT}. This set of fields contains the third higher dual
graviton $a_3:=A^{(3)}_{[2,2,2,1,1]}$ in the adjoint at level 4, and a number
of extra fields that are obtained by dualising the set independent fields
$\{a_2,c_1,\ldots,c_5\}$ from the previous level.
As before, looking at Table \ref{table:comparison},
we find that some fields are a perfect match and some are not. In fact,
for almost all of the extra fields introduced at this level, we have a surplus
of fields in the maximal off-shell description compared with the fields that
are available from the adjoint and $\ell_2$ representations. With the exception
of the rogue scalar field $d_9:=\widehat{O}^{(3)}_{[4,4]}$\,, the Young tableaux
for the extra fields perfectly match those of the spectrum of $\ell_2$ at
level 3. Rogue scalars like this one are found in the maximal off-shell
description at all odd levels of higher dualisation greater than or
equal to this level.

\paragraph{Optimal off-shell dualisation.}

In order to understand the differences between these higher
dualisation schemes, it is useful to give examples at low levels.
All three of them coincide at the first level of higher dualisation where the
dual graviton $A_{ab}$ is dualised to give $A^{ab,cd}$ and $\widehat{Z}^{abc,d}$\,.
We then have the choice of whether to dualise only the first higher dual graviton
or to dualise both fields. Unfortunately, we exceed the adjoint and $\ell_2$
representations of $A_1^{+++}$ if both of them are dualised. Optimal and minimal
off-shell dualisations coincide at this level. In fact, we do not have an exact
match with the fields coming from $A_1^{+++}$ because there is an extra $c_2$
field in the $A_1^{+++}$ tables that is not obtained in the optimal scheme.

At the next level of dualisation, we can choose any of the fields
in $\{a_1,c_1,c_2,c_3\}$ to dualise. It turns out that all of them may be dualised
to produce 16 new fields which are contained in the $A_1^{+++}$ representations.
However, we do not find a perfect match because there are six fields in the adjoint
and $\ell_2$ representations that cannot be obtained this way. In other words,
the representations of $A_1^{+++}$ contain slightly more fields than optimal
off-shell dualisation.

The adjoint and $\ell_2$ representations at the next level
contain 96 fields. If we dualise all of the fields at the previous level in the
optimal scheme, we obtain 69 fields. However, there are six $\mathbb{Y}[3,3,1,1]$
fields in the optimal scheme, whereas the adjoint and $\ell_2$ representations of
$A_1^{+++}$ only contain five. The set of fields that produce $\mathbb{Y}[3,3,1,1]$
upon dualisation is $\{4\times{}d_3,2\times{}d_8\}$\,, so optimal off-shell
dualisation is attained by choosing any one of these six fields not to dualise at
the previous level. This is important: optimal off-shell dualisation is, in general,
\emph{not} unique. Then again, we do not yet know what will happen at higher levels.
It is possible that these various pathways to optimal off-shell dualisation may
converge at higher levels. It would be interesting to draw the graph of optimal
pathways at higher levels and to study its topology.

\paragraph{More general statements.}

We have just observed what happens at low levels, but there is more to say.
It can easily be checked that the $n^\mathrm{th}$ higher dual graviton
$A^{(n)}\sim\mathbb{Y}[2,\ldots,2,1,1]$ corresponds to the $A_3$ weight $[0,n,2]$
with associated $A_1^{+++}$ root $(0,0,n,n+1)$ whose squared length is equal to 2.
That is, the $n^\mathrm{th}$ higher dual graviton appears in the adjoint at
level $n+1$. To calculate this, we have used equation (16.6.3) from \cite{West:2012vka}
while requiring that the Kac labels are non-negative. It can also be shown that the
$\ell_2$ representation of $A_1^{+++}$ contains the $\widehat{Y}^{(n)}$,
$\widehat{Z}^{(n)}$ and $\widehat{W}^{(n)}$ families of extra fields.
They correspond to the $A_3$ weights $[1,n-2,3]$, $[1,n-1,1]$
and $[0,n-2,2]$ with associated $A_1^{+++(2)}$ roots $(0,1,n-1,n,1)$, $(0,1,n,n,1)$
and $(1,2,n,n,1)$. However, even at low levels, this will turn out not to produce
the entire spectrum of $\ell_2$ and, indeed, less than half of the spectrum of
$\ell_2$ at level 3 is found if we only dualise the $A^{(n)}$ fields.

The $A_1^{+++}$ algebra has been shown to contain the minimal off-shell dualisation
of linearised gravity in four dimensions. Of course, extra fields may also be dualised
off-shell, but dualising too many of them leads to field multiplicities that exceed
those provided by the adjoint and $\ell_2$ representations of $A_1^{+++}$\,.
Maximal off-shell dualisation contains too many fields, but it is quite interesting
nonetheless. Despite some discrepancies in the multiplicities of
Table \ref{table:comparison}, the correct Young tableaux shapes appear in this
maximal scheme. More work is needed to fully understand the role of the rogue scalar
$\widehat{O}^{(3)}_{[4,4]}$ at the third level of higher dualisation, and the other
scalars at odd higher levels of dualisation.

The situation at the fourth level of maximal off-shell dualisation is more severe 
with fields that have a surplus as high as $+7$. However, ignoring multiplicities,
the Young tableaux at this level in the maximal off-shell description
perfectly matches the spectrum of $\ell_2$ at level 4.

In this section, we have identified a possible solution to the tricky problem of
mismatched multiplicities at each level: optimal off-shell dualisation.
In this scheme, one carefully chooses which extra fields to dualise so that, at each
level of higher dualisation, the set of extra fields is contained in the relevant 
representations of $A_1^{+++}$ with multiplicities that do not exceed those in the
`total' column in Table \ref{table:comparison}.

\section{The graviton tower action at low levels}
\label{section:repackage}

\noindent The off-shell dualisation procedure has the advantage that the extra fields 
contained in the $\ell_2$ representation of $A_1^{+++}$ are made explicit.
See, for example, the dual action \eqref{actionAZ} where the $\mathrm{GL}(4)$-irreducible 
field variables $A_{ab,cd}$ and $\widehat{Z}_{abc,d}$ are in direct contact with 
representations of $A_1^{+++}$\,, namely the adjoint and $\ell_2$ representations. 

In this section, at the first level of higher dualisation,
we show that the fields $A^{ab,cd}$ and $\widehat{Z}_{abc,d}$
may be repackaged into new fields: $\widetilde{A}_{ab}$ and
$\widetilde{A}^{ab,cd}$ with the respective 
symmetry types of $A^{(0)}=A_{ab}$ and $A^{(1)}=A^{ab,cd}$\,. 
We will show that the gauge transformation laws of these two fields 
are almost identical to those of the Fierz-Pauli field for
$\widetilde{A}_{ab}$ and the Labastida gauge field \cite{Labastida:1987kw} 
with symmetry type $\mathbb{Y}(3,1)=\mathbb{Y}[2,1,1]$ for
$\widetilde{A}^{ab,cd}$\,, with additional terms that entangle
the two gauge transformation laws.
In order to make contact with the Labastida formalism where
mixed-symmetry fields are given in the symmetric convention for Young tableaux,
we will also use this convention for the first higher dual graviton in this
section. However, it should be noted that this convention is not used in the
context of E\,-theory.

The equivalent formulation we will present for the action of the 
first higher dual graviton in terms of $\widetilde{A}^{ab,cd}$ 
and $\widetilde{A}_{ab}$
has the advantage of showing more explicitly 
the number of degrees of freedom through an on-shell duality 
relation between the gauge invariant curvature tensors of 
the two fields, as is usual in this context 
\cite{Hull:2000zn,Hull:2001iu,Bekaert:2002dt}. 

\paragraph{Change of variables.}

Recall that the fields appearing in the action \eqref{actionXZ} were 
$X_{ab;}{}^{ij}$ and $Z_{a;}{}^e$\,, the latter being the Hodge dual of 
$\widehat{Z}^{bcd,e}$\,, see \eqref{level1XA}.
The Hodge dual of the former field $X_{ab;}{}^{ef}$ is related to $A^{cd,ef}$
via $A^{ab,cd}\equiv\phi^{cd[a,b]}$ and 
$\phi^{abc,d}=\tfrac{3}{4}\,\varepsilon^{ijd(a}X_{ij;}{}^{bc)}$.
This field, the first higher dual graviton, transforms in the
$\mathrm{GL}(4)$-irreducible representation $\mathbb{Y}[2,1,1]=\mathbb{Y}(3,1)$\,.

From the independent field variables $X_{ab;}{}^{ij}$ and $Z_{a;}{}^e\,$ 
we introduce the two-form field
\begin{align}
\label{defU}
    U^{ab} := 
    \tfrac{1}{4}\,\varepsilon^{abcd}(X_{cd;e}{}^{e} - 4\, Z_{c;d})
    \equiv 
    -\tfrac{1}{2}\,\eta_{cd}\,\phi^{cd[a,\,b]} - \varepsilon^{abcd}\,Z_{c;d} 
    \equiv 
    -\tfrac{1}{2}\,A^{ab,c}{}_{c} + \widehat{Z}^{abc,}{}_{c} 
\end{align}
that transforms like 
\begin{align}
    \delta U^{ab} = 2\,\partial^{[a}\tau^{b]} + \varepsilon^{abcd}\,\partial_{c}\epsilon_{d}\;,\qquad 
    \tau^a := \lambda^{ab}{}_b - \mu_{b}{}^{b,\,a} \;,
\end{align}
while we recall from Section \ref{subsection:firsthigher} that 
the field $\phi^{abc,\,d}$ transforms like 
\begin{align}
    \delta \phi^{abc,\,d} =&\; 3\,\partial^d\lambda^{abc}
    - 3\,\partial^{(a}\lambda^{bc)d} + 3\,\partial^{(a}\mu^{bc),\,d}
    - \tfrac{3}{2}\,\eta^{(ab}\,\varepsilon^{c)dij}\,\partial_{i}\epsilon_j
    \;.
\end{align}
We define the $\mathbb{Y}(3,1)$-type gauge field 
\begin{align}
    \widetilde{\phi}^{abc,d} := \phi^{abc,d} + \tfrac{3}{4}\,\eta^{(ab}\,U^{c)d}
    \label{phitildetophiandU}
\end{align}
that transforms like 
\begin{align}
    \delta \widetilde\phi^{abc,d} =
    &\; 3\,\partial^d\widetilde{\lambda}^{abc}
    - 3\,\partial^{(a}\widetilde{\lambda}^{bc)d} 
    + 3\,\partial^{(a}\widetilde{\mu}^{bc),d}
    - \tfrac{3}{4}\,\eta^{(ab}\,\varepsilon^{c)def}\,\partial_e\epsilon_f\;,
\end{align}
where 
\begin{align}
    \widetilde{\lambda}^{abc} :=
    \lambda^{abc} - \tfrac{1}{4}\,\eta^{(ab}\,\tau^{c)}\;,
    \qquad 
    \widetilde{\mu}^{ab,c} := \mu^{ab,c} 
    + \tfrac{1}{6}\left(\eta^{ab}\tau^c - \eta^{c(a}\tau^{b)}\right)\;.
\end{align}
The advantage of this change of variable is that 
the newly defined gauge parameters 
$\widetilde{\lambda}^{abc}$ and $\widetilde{\mu}^{ab,c}$
have the same trace: 
\begin{align}
    \widetilde{\tau}^a := 
    \widetilde{\lambda}^{ab}{}_b - \widetilde{\mu}_{b}{}^{b,a} 
    = \tfrac{1}{2}\,\sigma^a - \tfrac{1}{2}\, \sigma^a\,\equiv\, 0 
    \;,\qquad 
    \sigma^a := \lambda^{ab}{}_b + \mu_{b}{}^{b,a}\;.
\end{align}
This also implies that, among the three linearly independent gauge fields
$\{\widetilde{\phi}^{abc,d}, U^{ab}, Z_{(a;b)}\}$, only $U^{ab}$ 
transforms with $\tau^a$\,. As a result, the dependence of the 
action $S[\widetilde{\phi}^{abc,d}, U^{ab}, Z_{(a;b)}]$ on $U^{ab}$ 
comes entirely through its field strength 
$H^{abc}(U):=3\,\partial^{[a}U^{bc]}\,$.

From \eqref{level1XphiZpsi-inverse} together with \eqref{phitildetophiandU} and 
\begin{align}
    \widetilde{X}_{ab;}{}^{cd} &= \tfrac{1}{2}\,\varepsilon_{abef}\,
    \widetilde{\phi}^{cde,f}\;,
    \qquad \widetilde{\phi}^{abc,d} = \tfrac{3}{4}\,\varepsilon^{ijd(a}\,\widetilde{X}_{ij;}{}^{bc)}\;,
    \label{xtildeandphitilde}
\end{align}
we obtain
\begin{align}
\widetilde{X}_{ab;}{}^{cd}
=X_{ab;}{}^{cd}+\tfrac{3}{8}\,\varepsilon_{abef}\eta^{(cd}U^{e)f} \;.
\end{align}
We now express the action \eqref{actionXZ} 
in terms of the independent fields $\widetilde{X}_{ab;}{}^{cd}\,$, 
$U^{ab}$ and $f_{ab}:=Z_{(a;b)}\,$:
\begin{align}\label{LagXUf}
    {\cal L}(\widetilde{X},U,f) = &  \;  
    \tfrac{7}{72}\,H_{abc}(U)\,H^{abc}(U) \,
    + \partial_{a}f_{bc}\,\partial^{a}f^{bc}\,
    - \tfrac{4}{3}\,\partial_{a}f^{ab}\,\partial_{c}f_{b}{}^{c}\,
    - \tfrac{5}{3}\,\partial_{a}f^{ab}\,
    \partial_{c}\widetilde{X}^{c}{}_{b;i}{}^{i}\,
    \nonumber \\
    &
    + 2\,\partial^{c}f^{ab}\,\partial^{d}\widetilde{X}_{da;bc}\,
    - \tfrac{1}{18}\,\varepsilon_{abcd}\,H^{abc}(U)\,\left(
    \partial_{i}\widetilde{X}^{id;e}{}_{e} +4\,\partial_{i}f^{id}\right) \,
    -\tfrac{1}{12}\,\partial_{a}\widetilde{X}^{ab;c}{}_{c}\,
    \partial^{i}\widetilde{X}_{ib;d}{}^{d}\,
    \nonumber \\
    &
    + \tfrac{1}{2}\,\partial_{a}\widetilde{X}^{ab;cd}\,
    \partial_{i}\widetilde{X}^{i}{}_{b;cd}\,
    +\tfrac{1}{2}\,\partial^{d}\widetilde{X}_{da;bi}\,
    \partial^{i}\widetilde{X}^{ab;c}{}_{c}\,
    + \tfrac{1}{16}\,\partial_{i}\widetilde{X}_{ab;c}{}^{c}\,
    \partial^{i}\widetilde{X}^{ab;d}{}_{d}\;,
\end{align}
where $H_{abc}(U):= 3\,\partial_{[a}U_{bc]}\,$.

Now we can dualise the field $U^{ab}$ into a scalar field $S$ by 
letting $H_{abc}$ be an independent field and creating a new parent 
Lagrangian ${\cal L}(\widetilde{X},H,f)$ with the additional term 
$\tfrac{1}{18}\,\varepsilon^{abcd}\,\partial_{a}S\,H_{bcd}$.
Solving the field equation for the auxiliary field $H_{abc}$
yields
\begin{align}
    H_{abc} = \tfrac{2}{7}\,\varepsilon_{abcd}
    \left( \partial_{i}\widetilde{X}^{id;e}{}_{e} 
+4\,\partial_{i}f^{id} + \partial^d S\,\right)\;.
\end{align}
Substituting this into the parent Lagrangian, we obtain the following dual Lagrangian which is 
given, up to a total derivative, by
\begin{align}
    {\cal L}(\widetilde{X},S,f)\,=\,& 
    \tfrac{1}{21}\,\partial_a S\,\partial^a S\,
    + \tfrac{8}{21}\,\partial_a S\,\partial_b f^{ab}
- \tfrac{4}{7}\,\partial_{a}f^{ab}\,\partial^{c}f_{cb}\,
+ \partial_{a}f_{bc}\,\partial^{a}f^{bc}\, 
\nonumber \\
& -\tfrac{1}{28}\,\partial_{a}\widetilde{X}^{ab;c}{}_{c}\,
\partial^{i}\widetilde{X}_{ib;d}{}^{d}\,
+ \tfrac{1}{2}\,\partial_{a}\widetilde{X}^{ab;cd}\,
\partial^{i}\widetilde{X}_{ib;cd}\,
+\tfrac{1}{2}\,\partial^{d}\widetilde{X}_{da;be}\,
\partial^{e}\widetilde{X}^{ab;c}{}_{c}\,
\nonumber\\
&  + \tfrac{1}{16}\,\partial_{d}\widetilde{X}_{ab;e}{}^{e}\,
\partial^{d}\widetilde{X}^{ab;c}{}_{c}\,
+ 2\,\partial^{c}f^{ab}\,\partial^{d}\widetilde{X}_{da;bc}\,
- \tfrac{9}{7}\,\partial_{a}f^{ab}\,
\partial^{c}\widetilde{X}_{cb;i}{}^{i}\;.
\end{align}
This action is invariant under
\begin{align}
    \delta \widetilde\phi_{abc,d} =&\;
    3\,\partial_d\widetilde{\lambda}_{abc}
    - 3\,\partial_{(a}\widetilde{\lambda}_{bc)d} 
    + 3\,\partial_{(a}\widetilde{\mu}_{bc),d}
    - \tfrac{3}{4}\,\eta_{(ab}\,\varepsilon_{c)dij}\,\partial^{i}\epsilon^j\;,\\
    \delta f_{ab} =&\; 
    \tfrac{1}{4}\left(\varepsilon_{a}{}^{cde}\,\partial_c\widetilde\mu_{bd,e}
    +\varepsilon_{b}{}^{cde}\,\partial_c\widetilde\mu_{ad,e}\right)
    +\partial_{(a}\epsilon_{b)}-\tfrac{1}{4}\,\eta_{ab}\,\partial_c\epsilon^c\;,\\
    \delta S =&\; -3\,\partial_a \epsilon^a\;.
\end{align}
%
%
Finally, we combine the scalar field $S$ with the traceless symmetric field $f_{ab}:=Z_{(a;b)}$ to get
\begin{align}
    \widetilde{A}_{ab}:= 2\,f_{ab} - \tfrac{1}{6}\,\eta_{ab}\,S
    \quad \Leftrightarrow \quad S = -\tfrac{3}{2}\,\widetilde{A}^a{}_a\;,\quad
    f_{ab} = \tfrac{1}{2}\,(\widetilde{A}_{ab} - \tfrac{1}{4}\,\eta_{ab}\,\widetilde{A}^c{}_c)\;.
\end{align}
We therefore obtain a new action $S[\widetilde{A}_{ab},\widetilde{A}^{ab,cd}]$
which takes the form
\begin{align}\label{firstdual}
S[\widetilde{A}_{ab}&,\widetilde{A}^{ab,cd}] =
-\tfrac{1}{2}\int \mathrm{d}^4x\,\Big[
-\tfrac{1}{2}\,\partial_a \widetilde{A}_{bc}\,\partial^a \widetilde{A}^{bc}
-\tfrac{3}{14}\,\partial_a \widetilde{A}_{b}{}^b\,\partial^a \widetilde{A}^{c}{}_c
+\tfrac{2}{7}\,\partial_a\widetilde{A}^{ab}\,\partial^c \widetilde{A}_{bc}
+\tfrac{3}{7}\,\partial_a \widetilde{A}^{ab}\,\partial_b \widetilde{A}_{c}{}^c\nonumber\\
&
\qquad\qquad\qquad\qquad+\tfrac{9}{7}\,\partial_a \widetilde{A}^{ab}\,\partial^c\widetilde{X}_{cb;d}{}^d
-2\,\partial^c \widetilde{A}^{ab}\,\partial^d \widetilde{X}_{da;bc}\\
&
+\tfrac{1}{14}\,\partial_a\widetilde{X}^{ab;c}{}_c\partial^d\widetilde{X}_{db;e}{}^e
-\tfrac{1}{8}\,\partial_d\widetilde{X}_{ab;e}{}^e\,\partial^d\widetilde{X}^{ab;c}{}_c
-\partial^d\widetilde{X}^{ab;c}{}_c\,\partial^e\widetilde{X}_{ea;bd}
-\partial_a\widetilde{X}^{ab;cd}\,\partial^e\widetilde{X}_{eb;cd}
\Big]\;,
\nonumber
\end{align}
where
\begin{equation}\label{repackagedAtildephitilde}
    \widetilde{A}^{ab,cd}=\widetilde{\phi}^{cd[a,b]}\;,
    \qquad
    \widetilde{\phi}^{abc,d}=-\tfrac{3}{2}\widetilde{A}^{d(a,bc)}
\end{equation}
and
\begin{equation}
    \widetilde{X}_{a_1a_2;}{}^{c_1c_2}\equiv\tfrac{1}{2}\,\varepsilon_{a_1a_2b_1b_2}\,\widetilde{\phi}^{c_1c_2b_1,b_2}\;,
    \qquad
    \widetilde{\phi}^{c_1c_2c_3,d}\equiv\tfrac{3}{4}\,\varepsilon^{ijd(c_1}\,\widetilde{X}_{ij;}{}^{c_2c_3)}\label{repackagedXtildephitilde}
\end{equation}
are understood throughout. This allows us to write the repackaged first higher
dual graviton $\widetilde{A}^{ab,cd}$ in a variety of useful ways.
For example, it was convenient to write the above action in terms
of $\widetilde{A}_{ab}$ and $\widetilde{X}_{ab;}{}^{cd}$\,. It is invariant under
the following intertwined gauge transformations:
\begin{align}
\delta \widetilde{A}_{ab} =&\; 2\,\partial_{(a}\epsilon_{b)} 
- \varepsilon_{cde(a}\partial^c\,\widetilde{\mu}_{b)}{}^{d,e}\;, 
\qquad\qquad \widetilde{\lambda}^{ab}{}_b \equiv   \widetilde{\mu}_{b}{}^{b,a}\;,
\label{gaugetransfoh}\\
\delta \widetilde\phi^{abc,d} =&\;
3\,\partial^d\widetilde{\lambda}^{abc}
- 3\,\partial^{(a}\widetilde{\lambda}^{bc)d} 
+ 3\,\partial^{(a}\widetilde{\mu}^{bc),d}
- \tfrac{3}{4}\,\eta^{(ab}\,\varepsilon^{c)dij}\,\partial_{i}\epsilon_j\;.
\label{gaugetransfophi1}
\end{align}
Trivially, one would need to make use of
\eqref{repackagedAtildephitilde} before checking gauge invariance under
\eqref{gaugetransfophi1}.
The $\widetilde{\lambda}^{abc}$ and $\widetilde{\mu}^{ab,c}$ parts of the
gauge transformations for $\widetilde{\phi}^{abc,d}$ coincide with the
Labstida gauge transformations for a gauge field of type $\mathbb{Y}(3,1)$.
In particular, the two gauge parameters are constrained to have equal trace.
The $\epsilon_a$ part of the gauge transformations for the (traceful)
symmetric rank-two tensor $\widetilde{A}_{ab}$ corresponds to linearised
diffeomorphisms. However, notice that $\widetilde{A}_{ab}$ also transforms
with the $\widetilde{\mu}^{ab,c}$ gauge parameter, and that
$\widetilde{\phi}^{abc,d}$ transforms with the gauge parameter $\epsilon_a\,$. 
As we have seen in \cite{Boulanger:2020yib} for higher dualisation of gauge fields, 
we find that the action contains fields that resemble the original dual graviton
$A_{ab}$ and the first higher dual graviton $\phi^{abc,d}$ with entangled gauge
transformation laws. By the construction of our dual action using the parent
action procedure, we know that the on-shell degrees of freedom are only
those of a single massless spin-2 field around four-dimensional Minkowksi
spacetime. Nevertheless, we will rederive this fact from the field equations.
It is clear that the dual action is more than just the sum of
the Fierz-Pauli and Labastida actions.

It is also important to remember that this repackaging approach
seeks to drastically redefine our fields for reasons that will become clear
towards the end of this section. As a result, gauge transformations
\eqref{gaugetransfoh} and \eqref{gaugetransfophi1} are not expected to resemble 
the gauge transformations for $A_{ab}$ and $A^{ab,cd}$ that were found in
Section \ref{subsection:A1+++gaugetransfo}
and Section \ref{subsection:firsthigher}.

\paragraph{Field equations.}

The equations of motion for the fields $\widetilde{A}_{ab}$ and $\widetilde\phi^{abc,d}$ are given by
\begin{equation}
    {\cal E}{[\widetilde{A}]}{}_{ab}\approx 0\qquad\mathrm{and}\qquad{\cal E}{[\widetilde{\phi}]}{}^{abc,d}\approx 0\;,
\end{equation}
where ${\cal E}{[\widetilde{A}]}{}_{ab}$ and ${\cal E}{[\widetilde{\phi}]}{}^{abc,d}$ 
are given by
\begin{align}
{\cal E}{[\widetilde{A}]}{}_{ab}:=\,&-\tfrac{1}{2}\Box \widetilde{A}_{ab}+\tfrac{3}{14}\partial_a\partial_b \widetilde{A}_c{}^c +\tfrac{2}{7}\partial^i\partial_{(a}\widetilde{A}_{b)i} -\tfrac{3}{14}\eta_{ab}
\left(\Box \widetilde{A}_c{}^c-\partial_i\partial_j\widetilde{A}^{ij}\right)\nonumber\\
&-\tfrac{9}{14}\partial^i\partial_{(a}\widetilde{X}_{b)ij}{}^j-\partial^i\partial^j\widetilde{X}_{i(ab)j}\;,\\
{\cal E}{[\widetilde{\phi}]}{}^{abc,d}
:=&\;\tfrac{1}{4}\Box\widetilde\phi^{abc,d}
-\tfrac{1}{4}\Box\widetilde\phi^{d(ab,c)}
-\tfrac{1}{4}\partial_i\partial^{d}\widetilde\phi^{abc,i}
+\tfrac{1}{4}\partial_i\partial^{(a}\widetilde\phi^{bc)d,i}
+\tfrac{1}{4}\partial^{d}\partial_i\widetilde\phi^{i(ab,c)}\nonumber\\
&
-\tfrac{1}{4}\partial_i\partial^{(a}\widetilde\phi^{bc)i,d}
-\tfrac{1}{8}\partial^{(a}\partial^b\widetilde\phi^{c)}{}_i{}^{i,d}
+\tfrac{1}{8}\partial^{(a}\partial^{b|}\widetilde\phi^{d}{}_i{}^{i,|c)}
+\tfrac{1}{8}\eta^{d(a}\partial_j\partial^b\widetilde\phi^{c)}{}_i{}^{i,j}
-\tfrac{1}{8}\eta^{(ab}\partial^{c)}\partial_i\widetilde\phi^{dij,}{}_j\nonumber\\
&
-\tfrac{1}{8}\eta^{d(a}\partial^{b|}\partial_j\widetilde\phi_i{}^{ij,|c)}
+\tfrac{1}{8}\eta^{(ab|}\partial_i\partial^d\widetilde\phi^{|c)ij,}{}_j
+\tfrac{1}{8}\eta^{(ab|}\partial_i\partial_j\widetilde\phi^{ijd,|c)}
-\tfrac{1}{8}\eta^{(ab}\partial_i\partial_j\widetilde\phi^{c)ij,d}\nonumber\\
&
-\tfrac{1}{7}\eta^{(ab|}\partial^{d}\partial_j\widetilde\phi_i{}^{ij,|c)}
-\tfrac{1}{7}\eta^{(ab}\partial^{c)}\partial_j\widetilde\phi^d{}_i{}^{i,j}
+\tfrac{15}{56}\eta^{(ab}\partial^{c)}\partial_j\widetilde\phi_i{}^{ij,d}
+\tfrac{1}{56}\eta^{(ab|}\partial_j\partial^d\widetilde\phi^{|c)}{}_i{}^{i,j}\nonumber\\
&
+\tfrac{5}{112}\eta^{(ab}\Box\widetilde\phi^{c)}{}_i{}^{i,d}
-\tfrac{5}{112}\eta^{(ab|}\Box\widetilde\phi^{d}{}_i{}^{i,|c)}
+\tfrac{1}{2}\varepsilon^{ijd(a}\partial_i\partial^{b}\widetilde{A}^{c)}{}_j+\tfrac{9}{28}\eta^{(ab}\varepsilon^{c)dij}\partial_i\partial^k\widetilde{A}_{jk}\;.
\end{align}
Alternatively, we may vary with respect to $\widetilde{X}_{ab;}{}^{cd}$ to find ${\cal E}{[\widetilde{X}]}{}_{ab;}{}^{cd}\approx 0$, where
\begin{align}
{\cal E}{[\widetilde{X}]}{}_{ab;}{}^{cd}:=\,&
-\partial_{[a}\partial^{(c}\widetilde{A}_{b]}{}^{d)}
+\tfrac{1}{4}\delta_{[a}{}^{(c}\Box{}\widetilde{A}_{b]}{}^{d)}
-\tfrac{1}{4}\delta_{[a}{}^{(c}\partial_{b]}\partial^{d)}\widetilde{A}_e{}^e
+\tfrac{1}{14}\delta_{[a}{}^{(c}\partial_{b]}\partial_{i}\widetilde{A}^{d)i}
-\tfrac{1}{14}\delta_{[a}{}^{(c}\partial^{d)}\partial^{i}\widetilde{A}_{b]i}\nonumber\\
&
+\tfrac{9}{14}\eta^{cd}\partial^{i}\partial_{[a}\widetilde{A}_{b]i}
+\partial^{i}\partial_{[a}\widetilde{X}_{b]i;}{}^{cd}
-\tfrac{1}{2}\partial^{(c}\partial_{[a}\widetilde{X}_{b]}{}^{d);i}{}_{i}
-\tfrac{1}{2}\eta^{cd}\partial^{i}\partial^{j}\widetilde{X}_{i[a;b]j}\nonumber\\
&
+\tfrac{1}{4}\delta_{[a|}{}^{(c|}\partial_{j}\partial^{i}\widetilde{X}_{i|b];}{}^{|d)j}
+\tfrac{1}{4}\delta_{[a|}{}^{(c|}\partial^{j}\partial_{i}\widetilde{X}^{i|d);}{}_{|b]j}
-\tfrac{1}{8}\eta^{cd}\Box\widetilde{X}_{ab;i}{}^{i}
-\tfrac{1}{14}\eta^{cd}\partial^{i}\partial_{[a}\widetilde{X}_{b]i;j}{}^{j}\nonumber\\
&
+\tfrac{5}{56}\delta_{[a}{}^{(c}\partial_{b]}\partial_{i}\widetilde{X}^{d)i;j}{}_{j}
+\tfrac{9}{56}\delta_{[a}{}^{(c}\partial^{d)}\partial^{i}\widetilde{X}^{b]i;j}{}_{j}\;.
\end{align}
They obey Noether identities associated with the gauge parameters. 
For $\epsilon_a$, we have
\begin{align}
    \partial^a{\cal E}{[\widetilde{A}]}{}_{ab} 
    - \tfrac{3}{8}\,\eta_{ij}\,\varepsilon_{klab}
    \partial^a{\cal E}{[\widetilde{\phi}]}{}^{ijk,l}\equiv 0\;.
\end{align}
In addition, associated with the traceless part of the 
$\widetilde{\mu}^{ab,c}$ gauge parameter, we find
\begin{align}
\Big(\varepsilon^{ijd(b}\partial_i{\cal E}{[\widetilde{A}]}{}_{j}{}^{c)}+3\partial_a{\cal E}{[\widetilde{\phi}]}{}^{abc,d}-3 \partial_a{\cal E}{[\widetilde{\phi}]}{}^{a(bc,d)}\Big)  - \mathrm{trace} \equiv 0\;,
\end{align}
where ``{trace}'' indicates the terms needed to remove the trace of the expression in the brackets. There are also Noether identities
related to the traceless part of the gauge parameter 
$\widetilde{\lambda}^{abc}$
and the shared trace of $\widetilde{\lambda}^{abc}$ and $\widetilde{\mu}^{ab,c}$ although we will not write them here. 

\paragraph{Gauge-invariant tensors.} Associated with the gauge transformations
\eqref{gaugetransfoh} and \eqref{gaugetransfophi1} for $\widetilde{A}_{ab}$ and $\widetilde{\phi}^{abc,d}$, respectively, 
we find the following gauge-invariant tensor with two derivatives:
\begin{align}
    K_{ma,nb} &:=
    4\, \partial_{[m}\partial_{[n}\widetilde{A}_{b]a]}  
    + \tfrac{10}{7}\, \eta_{[m[n}\partial_{b]}\partial^i \widetilde{A}_{a]i}
    + \tfrac{10}{7}\, \eta_{[n[m}\partial_{a]}\partial^i \widetilde{A}_{b]i}
    - \tfrac{20}{7}\, \eta_{[m[n}\partial_{b]}\partial_{a]} \widetilde{A}_e{}^e 
    \nonumber \\
    & - 4\, \partial_{[m|}\partial^i \widetilde{X}_{i[n;b]|a]}
    - 4 \,\partial_{[n|}\partial^i \widetilde{X}_{i[m;a]|b]}
    - \tfrac{5}{7}\, \eta_{[n[m}\partial_{a]}\partial^i \widetilde{X}_{b]i;j}{}^{j}
    - \tfrac{5}{7}\, \eta_{[m[n}\partial_{b]}\partial^i \widetilde{X}_{a]i;j}{}^{j}\;.
\end{align}
Notice that $I:=\Box \widetilde{A}_e{}^e - \partial^a\partial^b \widetilde{A}_{ab}\equiv -\tfrac{7}{16}\,K^{ab,}{}_{ab}\,$ is a gauge invariant invariant scalar.
This can be seen as resulting from the gauge transformation of
$\widetilde{V}_b := \partial^a[\widetilde{X}_{ab;c}{}^c + 2 (\widetilde{A}_{ab}-\eta_{ab}\widetilde{A}_c{}^c)]$\,:
\begin{align}
    \delta \widetilde{V}_b &= 7\,\partial^a\partial_{[a}\epsilon_{b]}\;.
\end{align}
We find that the left-hand side of the field equation for $\widetilde{A}_{ab}$ is related 
to the trace of $K_{ma,nb}\,$:
\begin{align}
-2\,{\cal E}{[\widetilde{A}]}{}_{ab} \equiv K_{ab} - \tfrac{1}{2}\,\eta_{ab}K =: G_{ab}\;,
\end{align}
where $K_{ab}:=\eta^{mn}K_{am,bn}$ and $K:=\eta^{ab}K_{ab}\,$.
Obviously, on-shell, we have $K_{ab}\approx 0\,$ which is to be compared 
with the field equation \eqref{eq(1.4.4)} in Section \ref{section:a1+++EOM}.
This is analogous to the Ricci-flat equation in linearised gravity.

We also have the following gauge-invariant quantity with three derivatives:
\begin{align}
G_{mn,pq;}{}^{d} := \;&4\,\varepsilon^{abcd} \,\partial_a\partial_{[m}\partial_{[p} \widetilde{\phi}_{q]n]b,c}
\nonumber \\
&-\tfrac{8}{7}\,\left( \eta_{[m[p}\partial_{q]}\partial_{n]}\widetilde{V}^d 
+\tfrac{1}{4}\,\big(\delta^{d}{}_{[m}\partial_{n]}\partial_{[p}\widetilde{V}_{q]}
+\delta^{d}{}_{[p}\partial_{q]}\partial_{[m}\widetilde{V}_{n]}\big) 
- \tfrac{1}{2}\,\eta_{m[p}\eta_{q]n}\partial^d\partial_a \widetilde{V}^a\right)\;
\end{align}
that possesses the algebraic symmetries of the Riemann tensor in its first
four indices and also satisfies two additional tracelessness constraints:
$G_{mn,pq;}{}^{m} \equiv G_{mn,pq;}{}^{p}\equiv 0$\,. These algebraic
constraints on the tensor $G_{mn,pq;}{}^{r}$\, imply that the dual tensor
$\widetilde{G}_{abc,mn,pq}:=\varepsilon_{abcd}\,G_{mn,pq;}{}^{d}$
is of $\mathrm{GL}(4)$-irreducible type $\mathbb{Y}[3,2,2]$\,.

\paragraph{The Bianchi identity.} We find that $\partial_{[a}K_{bc],de}$ is expressed in terms of $\widetilde{A}_{ab}$ and $\widetilde{\phi}^{abc,d}$ as
\begin{equation}    \partial_{[a}K_{bc],de}=4\,\partial^i\partial_{[d}\partial_{[a|}\widetilde{X}_{i|b;c]e]}
+\tfrac{5}{7}\eta_{[a[d}\partial^i\partial_{e]}\partial_b\widetilde{X}_{c]i;j}{}^{j}
-\tfrac{10}{7}\eta_{[a[d}\partial^i\partial_{e]}\partial_b\widetilde{A}_{c]i}\;.
\end{equation}
It is possible to write this in terms of the left-hand-side of the field
equation for $\widetilde{X}_{ab;}{}^{cd}$\,:
\begin{align}
\partial_{[a}K_{bc],de}\equiv\;
&
8\,\partial_{[a}{\cal E}{[\widetilde{X}]}{}_{b[d;e]c]}
-
\eta_{[a[d}\partial_{e]}{\cal E}{[\widetilde{X}]}{}_{bc];i}{}^{i}
+\tfrac{4}{3}\,\eta_{[d[a}\partial^{i}{\cal E}{\widetilde{X}}{}_{bc];e]i}
+\tfrac{8}{3}\,\eta_{[a[d}\partial^{i}{\cal E}{[\widetilde{X}]}{}_{e]b;c]i}\;.
\end{align}
Therefore, on-shell, we find the following relation that will be instrumental 
in showing that the degrees of freedom are those of a single graviton:
\begin{equation}
	\partial_{[a}K_{bc],de}\approx0\;.
	\label{onshellBianchi}
\end{equation}

\paragraph{On-shell duality relation.}
We find that the gauge-invariant tensor $G_{ab,cd;}{}^e$ is related to $K_{ab,cd}$ and the left-hand-sides of the equations of motion in the following way:
\begin{align}
{G_{ab,cd;}}^e\equiv&-\partial^eK_{ab,cd}\nonumber\\
&
+\eta_{[a[c}\partial_{d]}{\cal E}{[\widetilde{A}]}{}_{b]}{}^e
+\eta_{[c[a}\partial_{b]}{\cal E}{[\widetilde{A}]}{}_{d]}{}^e
-\delta_{[a}{}^{e}\partial_{[c}{\cal E}{[\widetilde{A}]}{}_{d]b]}
-\delta_{[c}{}^{e}\partial_{[a}{\cal E}{[\widetilde{A}]}{}_{b]d]}
+\eta_{a[c}\eta_{d]b}\partial^{e}{\cal E}{[\widetilde{A}]}{}_{i}{}^{i}\nonumber\\
&
+8\,\partial_{[a}{\cal E}{[\widetilde{X}]}{}^{e}{}_{[c;d]b]}
+8\,\partial_{[c}{\cal E}{[\widetilde{X}]}{}^{e}{}_{[a;b]d]}
+2\,\eta_{[a[c}\partial_{d]}{\cal E}{[\widetilde{X}]}{}_{b]}{}^{e;i}{}_{i}
+2\,\eta_{[c[a}\partial_{b]}{\cal E}{[\widetilde{X}]}{}_{d]}{}^{e;i}{}_{i}\;.
\end{align}
Consequently, on-shell, we have the following duality relation:
\begin{equation}
    \widetilde{G}_{a[3],b[2],c[2]} \approx - \;\varepsilon_{a[3]d}\,
    \partial^d K_{b[2],c[2]}
    \qquad\Leftrightarrow\qquad
    G_{mn,pq;}{}^r \approx - \;\partial^r K_{mn,pq}\;.
    \label{twistedduality}
\end{equation}

These equations are important in several respects. 
The tensors $K_{ab,cd}$ and $\widetilde{G}_{abc,de,fg}$ 
can be called the field strengths for the repackaged dual graviton
and first higher dual graviton, resepctively.
Indeed, they do not vanish on-shell and they are gauge invariant. 
The duality relation \eqref{twistedduality} sets equal the two 
curvature tensors, on-shell, thereby showing that the 
physical degrees of freedom carried by the field $\widetilde{A}_{ab}$ 
are also contained in $\widetilde{\phi}_{abc,d}\,$.
There is no doubling of the degrees of freedom. 
Secondly, from \eqref{twistedduality} and the on-shell 
Bianchi identity \eqref{onshellBianchi}, we find
\begin{equation}
    \widetilde{G}^{ab}{}_{c,ab,de}\approx 0\;,
\end{equation}
which is exactly the form of the field equation \eqref{eq(1.4.10)} that we derived
for the first higher dual graviton in Section \ref{section:a1+++EOM}.
Finally, by taking the trace of the duality relation \eqref{twistedduality}
on the indices $b_2$ and $c_2\,$ and using the Ricci-flat equation 
$K_{ab}\approx 0$ that we derived above, we find
\begin{equation}
    \widetilde{G}_{abc,de,}{}^{d}{}_f \approx 0\;.
\end{equation}
This field equation completes those found in Section \ref{section:a1+++EOM}.

With these field equations, we have found a strong parallel with the 
analogous equations derived in Section \ref{section:a1+++EOM}. 
However, since no action principle was considered in that section, 
each field strength was a function of a single field. 
Instead, in the off-shell formulation found in the present section
that requires the extra field $\widehat{Z}^{abc,d}$ to be repackaged,
equations necessarily entangle both fields due to the nature of the 
gauge transformation laws.

We conjecture that this dualisation and repackaging procedure creates an 
increasingly tall tower of new repackaged dual gravitons whose gauge transformation 
laws are intertwined. In particular, for a given tower with highest level $N$, 
the field $\widetilde{\phi}^{(n)}$ at level $n\leq{}N$ should transform as a Labastida 
gauge field of symmetry type $\mathbb{Y}[2,\ldots,2,1,1]\,$. 
Its gauge transformation law 
should contain terms that entangle it with the repackaged fields at every level lower 
than $n\,$. Moreover, if $n<N\,$, then its gauge transformation law will also be 
entangled with that of the repackaged field at level $n+1\,$. It may even be possible 
to redefine fields so that the repackaged dual graviton at level $n\,$ is entangled 
only with those at level $n+1$ and $n-1\,$.

For the graviton tower action $S[\widetilde{A}_{ab},\widetilde{A}^{ab,cd}]$ 
in \eqref{firstdual}, the extra field $\widetilde{Z}^{abc,d}$ was completely hidden
by the specific field and gauge parameter redefinitions used to construct 
$\widetilde{A}_{ab}$\,. 
However, this may only be possible at low levels, so we cannot yet exclude the 
possibility that some extra fields may still be present in the graviton tower 
actions at higher levels.

\section{Conclusion}

In this paper, we started to make precise connections between the non-linear
realisation based on $A_1^{+++}$ 
\cite{Lambert:2006he,Tumanov:2014pfa,Pettit:2017zgx,Glennon:2020qpt,Glennon:2020uov} 
and the off-shell dualisation programme for pure gravity in four dimensions
\cite{Boulanger:2012df}. The non-linear realisation contains an 
infinite number of dualisations of gravity.
It consists of an infinite set of duality relations, the first of which involves
only the graviton and the dual graviton. This relation was worked out at the
full non-linear level in \cite{Glennon:2020qpt,Glennon:2020uov}.
In Section 2, we have used the non-linear realisation to work out
the linearised equations of motion for the first higher dual graviton.

While on the other hand, in \cite{Boulanger:2012df} 
it was shown that pure linearised gravity could be 
described by any member of an infinite family of action principles, 
each involving more and more fields.
Some of these fields were shown in \cite{Boulanger:2012df} to 
have a direct connection with the adjoint representation of 
the very-extended $A_{D-3}^{+++}$ algebra, 
while other fields received no interpretation at that time.

In the present paper, where we focus on $D=4$ for the sake of concreteness,
we showed that the aforementioned fields are all associated with
generators in the $\ell_2$ representation of $A_1^{+++}$ in the sense 
that there exist generators in $\ell_2$ that have the same $\mathrm{GL}(4)$
types. We have carried out this match up to level four and, while there is a 
striking agreement at low levels, some of the multiplicities differ for the extra 
fields.

We also constructed, at the level of the first higher dual graviton, 
a new action principle featuring two fields $\widetilde{A}_{ab}$ and
$\widetilde{A}_{ab,cd}$ with the $\mathrm{GL}(4)$ symmetry types
$\mathbb{Y}[1,1]$ and $\mathbb{Y}[2,1,1]$ of the dual graviton and the
first higher dual graviton, respectively. The gauge transformations of
these two fields are those of the dual graviton and the corresponding
$\ytableaushort{\null\null\null,\null}$ Labastida field, along with extra
terms that entangle the two fields. Remarkably, the field equations can be
obtained from a duality relation between the gauge invariant curvatures
of these repackaged fields, which further demonstrates that our original
action only propagates a single graviton. That the field equations can be
encapsulated in a set of duality relations is in full agreement with the 
method of obtaining the field equations in the non-linear realisation 
of $A_1^{+++}\ltimes\ell_1\,$.

In a future work in preparation, we will extend our analysis to 
pure gravity in five dimensions where the relevant algebra is 
$A_2^{+++}\,$. We will also consider pure gravity and 
the bosonic sector of maximal supergravity in eleven dimensions.
It is well-known that the relevant Kac-Moody algebras for 
these theories are $A_8^{+++}$ and $E_{11}\,$, respectively.
There, we will also show how their $\ell_2$ representations 
are related to the set of off-shell fields entering higher dual 
action principles. It will be important to
modify the coset space used to construct the non-linear realisation
for these algebras in order to incorporate $\ell_2$\,.
Consequently, this will account for the extra fields that were
thought to be missing from E\,-theory until now.

Finally, it would be interesting to make a contact with
\cite{Bossard:2021ebg} where the importance of the $\ell_2$
representation of $E_{11}$ was noticed in a similar context.
It is not yet clear to us that there is a connection since
their equations of motion are obtained from the $E_{11}$ pseudo\,-Lagrangian
by a variational principle supplemented by extra duality relations that are
not derived by variation. More specifically, variations with respect to
constrained fields (which carry a section constraint index) vanish only when
these extra duality relations are imposed. In contrast, our off-shell
dualisation approach produces equations of motion and duality relations that
are all obtained by varying dual actions. Nothing external needs to be
imposed here.
Another line of research is to investigate the possible 
non-linear extensions of the higher dual actions considered here.

\section*{Acknowledgements}
N.B. and J.A.O. wish to thank Andrea Campoleoni and Victor Lekeu 
for discussions.
We have performed and/or checked several computations 
with the package xTras 
\cite{Nutma:2013zea} of the suite of Mathematica packages xAct.
The $A_1^{+++}$ representations in this paper were produced using
the programme SimpLie \cite{Bergshoeff:2007qi}.
The work of N.B. is partially supported by the F.R.S.-FNRS PDR grant 
``Fundamental issues in extended gravity'' No. T.0022.19.
The work of J.A.O. is supported by the European Research Council (ERC) under the European 
Union's Horizon 2020 research and innovation programme (grant agreement No. 101002551).
P.W. would like to thank the STFC, grant numbers ST/P000258/1 
and ST/T000759/1, for support.

\clearpage
\appendix

\section{Young tableaux in the symmetric convention}\label{appendix-phipsi}

During the construction of the first higher dual action for gravity, we might want
to express the irreducible field content $\{A^{ab,cd}\,,\widehat{Z}^{abc,d}\}$ of 
\eqref{actionAZ} in terms of fields with blocks of symmetric indices,
corresponding to the manifestly
symmetric convention for Young tableaux. Recall that 
$\{A^{ab,cd}\,,\widehat{Z}^{abc,d}\}$ and
$\{X_{ab;}{}^{cd}\,,\widehat{Z}_{a;}{}^{b}\}$ are related by \eqref{level1XA} and \eqref{level1XA-inverse} as follows:
\begin{gather}
    A^{ab,cd} := -\tfrac{1}{2}\,\varepsilon^{abij}\,X_{ij;}{}^{cd}\;,\qquad
    \widehat{Z}^{abc,d} := \varepsilon^{abce}\,Z_{e;}{}^d\;,\\
    X_{ab;}{}^{cd}=\tfrac{1}{2}\,\varepsilon_{abij}\,A^{ij,cd}\;,
    \qquad
    Z_{a;}{}^{e}=\tfrac{1}{6}\,\varepsilon_{abcd}\,\widehat{Z}^{bcd,e}\;.
\end{gather}
We can now introduce equivalent fields in the symmetric convention:
\begin{equation}
    \phi^{c_1c_2c_3,d} := \tfrac{3}{4}\,\varepsilon^{ijd(c_1}\,X_{ij;}{}^{c_2c_3)}\;,\qquad
    \psi^{c_1c_2,d,e} := \tfrac{1}{2}\,\varepsilon^{aed(c_1}
    {Z}_{a;}{}^{c_2)}\;.
    \label{level1XphiZpsi-inverse}
\end{equation}
Inverse relations are given by
\begin{equation}
    X_{a_1a_2;}{}^{c_1c_2} = 
    \tfrac{1}{2}\,\varepsilon_{a_1a_2b_1b_2}\,\phi^{b_1c_1c_2,b_2}\;,
    \qquad  {Z}_{a;}{}^{c} = \tfrac{1}{2}\,\varepsilon_{ab_1b_2b_3}\psi^{b_1c,b_2,b_3}\;.
    \label{level1XphiZpsi}
\end{equation}
These $\mathrm{GL}(4)$-irreducible fields satisfy over-symmetrisation constraints:
\begin{align}
    \phi^{abc,d} &= \phi^{(abc),d}\;, &\phi^{(abc,d)}\equiv 0\;,\\
    \psi^{ab,c,d} &= \psi^{(ab),c,d}\;, &\psi^{(ab,c),d}\equiv\psi^{(ab|,c,|d)}\equiv\psi^{ab,(c,d)}\equiv 0\;.
\end{align}
This is an opportunity to summarise and exemplify the two equivalent conventions 
for Young tableaux of $\mathrm{GL}(D)$\,. A finite-dimensional irreducible representation
of $\mathrm{GL}(D)$\, may be described by a tensor field with groups of manifestly
symmetric indices, each group corresponding to a row on the Young tableau
associated with it, such that the tensor satisfies over-symmetrisation identities.
Alternatively, we may choose to describe the same finite-dimensional irreducible
representation of $\mathrm{GL}(D)$\, by a tensor with groups of manifestly antisymmetric
indices, each group corresponding to a column on the Young tableau, 
such that the corresponding tensor satisfies over-antisymmetrisation identities. 
The $\mathrm{GL}(4)$ symmetries of the fields introduced so far 
are depicted by the following Young tableaux:
\begin{equation}
    \ytableausetup{smalltableaux}
    \phi^{abc,d}~\sim~ \begin{ytableau}
    a & b & c \\
    d 
    \end{ytableau}\;\sim\; A^{ad,bc}\;,
    \qquad
    \psi^{ab,c,d}~\sim~
    \begin{ytableau}
    a & b \\
    c \\
    d
    \end{ytableau}\;\;\sim\; \widehat{Z}^{acd,b}\;.
\end{equation}
The relation between the two conventions for Young tableaux is given by 
\begin{align}
    A^{ab,cd} \equiv \phi^{cd[a,b]}\;,
    \qquad
    \widehat{Z}^{abc,d} \equiv 3\,\psi^{d[a,b,c]}\;,\label{level1newXphiZpsi}
\end{align}
with inverse relations
\begin{align}
    \phi^{abc,d} \equiv -\tfrac{3}{2}\,A^{d(a,bc)}\;,
    \qquad
    \psi^{ab,c,d} \equiv \tfrac{1}{2}\,\widehat{Z}^{cd(a,b)}\;.
    \label{level1newXphiZpsi-inverse}
\end{align}
These relations describe nothing more than a change of basis between
irreducible tensor fields in the manifestly symmetric and antisymmetric
conventions for Young tableaux.

The second level of higher dualisation involves a reducible field
$D_{ab;}{}^{c_1c_2c_3,d}$ which must be decomposed into traceless
components as in \eqref{level2projection}. These components then need
to be Hodge dualised on their first blocks of indices in the same way
that $X_{ab;}{}^{cd}$ and $Z_{a;}{}^{b}$ were.
As before, this will create $\mathrm{GL}(4)$-irreducible fields with symmetric blocks
of indices. Their symmetry types are $\mathbb{Y}(4,2)$, $\mathbb{Y}(4,1,1)$,
$\mathbb{Y}(3,2,1)$ and $\mathbb{Y}(3,1,1,1)$\,:
\begin{align}
    \phi^{c_1c_2c_3c_4,d_1d_2}&=\tfrac{4}{5}\,
    \varepsilon^{a_1a_2(d_1(c_1}X_{a_1a_2;}{}^{c_2c_3c_4),d_2)}\;,\\
    \psi_{_{(Y)}}^{c_1c_2c_3c_4,d,e}&=\tfrac{2}{3}\,\varepsilon^{aed(c_1}Y_{a;}{}^{c_2c_3c_4)}\;,\\
    \psi_{_{(Z)}}^{c_1c_2c_3,d_1d_2,e}&=\tfrac{4}{5}\,\varepsilon^{ae(d_1(c_1}Z_{a;}{}^{c_2c_3),d_2)}\;,\\
    \psi_{_{(W)}}^{c_1c_2c_3,d,e,f}&=-\tfrac{1}{6}\,\varepsilon^{fed(c_1}W^{c_2c_3)}\;,
\end{align}
Inverse relations are given by
\begin{align}
    X_{a_1a_2;}{}^{c_1c_2c_3,d}&=\tfrac{1}{2}\varepsilon_{a_1a_2b_1b_2}\phi^{b_1c_1c_2c_3,b_2d}\;,\label{symlevel2X}\\
    Y_{a;}{}^{c_1c_2c_3}&=\tfrac{1}{2}\varepsilon_{ab_1b_2b_3}\psi_{_{(Y)}}^{b_1c_1c_2c_3,b_2,b_3}\;,\label{symlevel2Y}\\
    Z_{a;}{}^{c_1c_2,d}&=\tfrac{1}{2}\varepsilon_{ab_1b_2b_3}\psi_{_{(Z)}}^{b_1c_1c_2,b_2d,b_3}\;,\label{symlevel2Z}\\
    W^{c_1c_2}&=\tfrac{1}{2}\varepsilon_{b_1b_2b_3b_4}\psi_{_{(W)}}^{b_1c_1c_2,b_2,b_3,b_4}\;.\label{symlevel2W}
\end{align}
The symmetries of the fields on the right-hand-sides of \eqref{symlevel2X}--\eqref{symlevel2W} are depicted as
\begin{equation}
    \ytableausetup{notabloids}
    \begin{ytableau}
    b_1 & c_1 & c_2 & c_3 \\
    b_2 & d
    \end{ytableau}
    \quad,\qquad
    \ytableausetup{notabloids}
    \begin{ytableau}
    b_1 & c_1 & c_2 & c_3 \\
    b_2 \\
    b_3
    \end{ytableau}
    \quad,\qquad
    \ytableausetup{notabloids}
    \begin{ytableau}
    b_1 & c_1 & c_2 \\
    b_2 & d \\
    b_3
    \end{ytableau}
    \quad,\qquad
    \ytableausetup{notabloids}
    \begin{ytableau}
    b_1 & c_1 & c_2 \\
    b_2 \\
    b_3 \\
    b_4
    \end{ytableau}\quad.
\end{equation}

Generators of $A_1^{+++}$ are usually written with antisymmetric indices. 
To match that, we take the irreducible fields in the symmetric convention, 
namely $\phi^{c_1c_2c_3c_4,d_1d_2}$, $\psi_{_Y}^{c_1c_2c_3c_4,d,e}$, 
$\psi_{_Z}^{c_1c_2c_3,d_1d_2,e}$ and $\psi_{_W}^{c_1c_2c_3,d,e,f}\,$,  
and use them to construct irreducible fields with antisymmetric blocks of
indices which obey over-antisymmetrisation identities. They are defined by
\begin{align}
    A^{a_1a_2,b_1b_2,cd} &:= \phi^{cd[a_1[b_1,b_2]a_1]}\label{antisymlevel2X}\;,\\
    \widehat{Y}^{a_1a_2a_3,c_1,c_2,c_3} &:= \psi_{_{(Y)}}^{c_1c_2c_3[a_1,a_2,a_3]}\label{antisymlevel2Y}\;,\\
    \widehat{Z}^{a_1a_2a_3,b_1b_2,c} &:= \psi_{_{(Z)}}^{c[a_1[b_1,b_2]a_2,a_3]}\label{antisymlevel2Z}\;,\\
    \widehat{W}^{a_1a_2a_3a_4,c_1,c_2} &:= \psi_{_{(W)}}^{c_1c_2[a_1,a_2,a_3,a_4]}\label{antisymlevel2W}\;,
\end{align}
with inverse relations
\begin{align}
    \phi^{c_1c_2c_3c_4,d_1d_2} &= -\tfrac{12}{5}\,A^{(c_1|(d_1,d_2)|c_2,c_3,c_4)}\label{antisymlevel2Xinverse}\;,\\
    \psi_{_{(Y)}}^{c_1c_2c_3c_4,d,e} &= 2\,\widehat{Y}^{de(c_1,c_2,c_3,c_4)}\label{antisymlevel2Yinverse}\;,\\
    \psi_{_{(Z)}}^{c_1c_2c_3,d_1d_2,e} &= \tfrac{16}{5}\,\widehat{Z}^{(c_1|e(d_1,d_2)|c_2,c_3)}\label{antisymlevel2Zinverse}\;,\\
    \psi_{_{(W)}}^{c_1c_2c_3,d,e,f} &= 2\,\widehat{W}^{fed(c_1,c_2,c_3)}\label{antisymlevel2Winverse}\;.
\end{align}

\newpage
\section{Representations of $A_1^{+++}$ at the next level}\label{appendix-l2higher}

\begin{table}[h]
\caption{The adjoint representation of $A_1^{+++}$ at level five.}
\centering
\begin{tabular}{|c|c|c|c|c|c|}\hline
$l$&$A_{3}$ weight&$A_1^{+++}$ root $\alpha$&$\alpha^2$&mult.&field\\\hline\hline
$5$&$[0,1,0]$&$(2,4,7,5)$&$-24$&$1$&$A_{[4,4,2]}$\\
$5$&$[2,0,0]$&$(1,4,7,5)$&$-22$&$1$&$A_{[4,3,3]}$\\
$5$&$[0,0,2]$&$(2,4,6,5)$&$-22$&$2$&$A_{[4,4,1,1]}$\\
$5$&$[1,1,1]$&$(1,3,6,5)$&$-20$&$5$&$A_{[4,3,2,1]}$\\
$5$&$[3,0,1]$&$(0,3,6,5)$&$-16$&$2$&$A_{[3,3,3,1]}$\\
$5$&$[1,0,3]$&$(1,3,5,5)$&$-16$&$3$&$A_{[4,3,1,1,1]}$\\
$5$&$[0,3,0]$&$(1,2,6,5)$&$-16$&$2$&$A_{[4,2,2,2]}$\\
$5$&$[2,2,0]$&$(0,2,6,5)$&$-14$&$3$&$A_{[3,3,2,2]}$\\
$5$&$[0,2,2]$&$(1,2,5,5)$&$-14$&$3$&$A_{[4,2,2,1,1]}$\\
$5$&$[2,1,2]$&$(0,2,5,5)$&$-12$&$4$&$A_{[3,3,2,1,1]}$\\
$5$&$[0,1,4]$&$(1,2,4,5)$&$-8$&$1$&$A_{[4,2,1,1,1,1]}$\\
$5$&$[2,0,4]$&$(0,2,4,5)$&$-6$&$2$&$A_{[3,3,1,1,1,1]}$\\
$5$&$[1,3,1]$&$(0,1,5,5)$&$-8$&$3$&$A_{[3,2,2,2,1]}$\\
$5$&$[1,2,3]$&$(0,1,4,5)$&$-4$&$2$&$A_{[3,2,2,1,1,1]}$\\
$5$&$[0,4,2]$&$(0,0,4,5)$&$2$&$1$&$A_{[2,2,2,2,1,1]}$\\\hline
\end{tabular}
\label{A1+++-adj-level5}
\vspace{-0.5cm}
\end{table}


\begin{table}[h]
\caption{The $\ell_2$ representation of $A_1^{+++}$ at level four.}
\centering
\begin{tabular}{|c|c|c|c|c|c|}\hline
$l$&$A_{3}$ weight&$A_1^{+++(2)}$ root $\alpha$&$\alpha^2$&mult.&field\\\hline\hline
$4$&$[0,1,0]$&$(2,4,6,4,1)$&$-22$&$3$&$A_{[4,4,2]}$\\
$4$&$[2,0,0]$&$(1,4,6,4,1)$&$-20$&$3$&$A_{[4,3,3]}$\\
$4$&$[0,0,2]$&$(2,4,5,4,1)$&$-20$&$3$&$A_{[4,4,1,1]}$\\
$4$&$[1,1,1]$&$(1,3,5,4,1)$&$-18$&$12$&$A_{[4,3,2,1]}$\\
$4$&$[3,0,1]$&$(0,3,5,4,1)$&$-14$&$3$&$A_{[3,3,3,1]}$\\
$4$&$[1,0,3]$&$(1,3,4,4,1)$&$-14$&$7$&$A_{[4,3,1,1,1]}$\\
$4$&$[0,3,0]$&$(1,2,5,4,1)$&$-14$&$6$&$A_{[4,2,2,2]}$\\
$4$&$[2,2,0]$&$(0,2,5,4,1)$&$-12$&$3$&$A_{[3,3,2,2]}$\\
$4$&$[0,2,2]$&$(1,2,4,4,1)$&$-12$&$9$&$A_{[4,2,2,1,1]}$\\
$4$&$[2,1,2]$&$(0,2,4,4,1)$&$-10$&$5$&$A_{[3,3,2,1,1]}$\\
$4$&$[0,1,4]$&$(1,2,3,4,1)$&$-6$&$4$&$A_{[4,2,1,1,1,1]}$\\
$4$&$[2,0,4]$&$(0,2,3,4,1)$&$-4$&$1$&$A_{[3,3,1,1,1,1]}$\\
$4$&$[1,3,1]$&$(0,1,4,4,1)$&$-6$&$3$&$A_{[3,2,2,2,1]}$\\
$4$&$[1,2,3]$&$(0,1,3,4,1)$&$-2$&$2$&$A_{[3,2,2,1,1,1]}$\\\hline
\end{tabular}
\label{A1+++-l2-level45}
\vspace{-5.0cm}
\end{table}

\newpage
\providecommand{\href}[2]{#2}\begingroup\raggedright\endgroup


\end{document}